\documentclass{sig-alternate}
\pdfoutput=1 
\usepackage{algorithm}
\usepackage{algpseudocode}
\usepackage{amsmath}
\usepackage{appendix}
\usepackage{balance}
\usepackage{color}
\usepackage{comment}
\usepackage{epstopdf}
\usepackage{graphicx}
\usepackage{listings}
\usepackage{mathtools}
\usepackage{microtype}
\usepackage{multirow}
\usepackage{pdfpages}
\usepackage{subfig}
\usepackage{tabularx}
\usepackage{times}
\usepackage{url}
\usepackage{xcolor}
\usepackage{xspace}
\usepackage{bm}
\usepackage{mdwlist}

\newcommand{\subparagraph}{}
\usepackage{titlesec}

\titlespacing{\section}{0ex}{0.8ex}{0ex}
\titlespacing{\subsection}{0ex}{0.8ex}{0ex}
\titlespacing{\subsubsection}{0ex}{0.7ex}{0ex}
\algrenewcommand\alglinenumber[1]{\scriptsize #1:}

\makeatletter
\renewcommand*{\ALG@name}{Investing Rule}
\makeatother

\usepackage{breqn}
\newcommand{\ainv}{$\alpha$-investing }
\newcommand{\sfdr}{Sequential-FDR }

\newcommand{\Ex}[1]{E\left[#1\right]}

\newcommand{\pval}{$p$-value }
\newcommand{\pvals}{$p$-values }

\newcommand{\system}{{\sc Aware}}
\newcommand{\Chi}{\mathcal{X}}

\newtheorem{theorem}{Theorem}

\definecolor{dark-gray}{gray}{0.2}

\newcounter{tempEquationCounter}
\newcounter{thisEquationNumber}


\newcounter{parentnumber}

\begin{document}

\title{Controlling False Discoveries During\\ Interactive Data Exploration}

\numberofauthors{1}
\author{
\alignauthor
\vspace*{-30pt}
\begin{tabular}{cccc}
Zheguang Zhao &  Lorenzo De Stefani & Emanuel Zgraggen & Carsten Binnig  \\
 & Eli Upfal & Tim Kraska\\
\end{tabular}\\
\vspace{1.5mm}
\affaddr{Department of Computer Science, Brown University}\\
\vspace{0.75mm}
\{firstname\_lastname\}@brown.edu
}

\date{}
\maketitle

\begin{abstract}
Recent tools for interactive data exploration significantly increase the chance that users make false discoveries.
The crux is that these tools implicitly allow the user to test a large body of different hypotheses with just a few clicks thus incurring in the issue commonly known in statistics as the ``\emph{multiple hypothesis testing error}''.
In this paper, we propose solutions to integrate multiple hypothesis testing control into interactive data exploration tools.
A key insight is that existing methods for controlling the false discovery rate (such as FDR) are not directly applicable for interactive data exploration.
We therefore discuss a set of new control procedures that are better suited and integrated them in our system called \system{}.
By means of extensive experiments using both real-world and synthetic data sets we demonstrate how \system{} can help experts and novice users alike to efficiently control false discoveries. 
\end{abstract}

\section{Introduction}
\label{sec:intro}
``Beer is good for you: study finds that suds contain anti-viral powers'' [DailyNews 10/12].
``Secret to Winning a Nobel Prize? Eat More Chocolate''  [Time, 10/12]. 
``Scientists find the secret of longer life for men (the bad news: Castration is the key)'' [Daily Mail UK, 09/12]. 
``A new study shows that drinking a glass of  wine is just as good as spending an hour at the gym'' [Fox News, 02/15].

There has been an explosion of data-driven discoveries like the ones mentioned above. While several of these are legitimate, there is an increasing concern that a large amount of current published research findings are false \cite{Ioannidis_2005}.
The reasons behind this trend are manifold. 

In this paper we make the case that the rise of interactive data exploration (IDE) tools has the potential to worsen this situation further. 
Commercial tools like Tableau or research prototypes like Vizdom \cite{vizdom}, Dice \cite{dice} or imMens \cite{immens}, aim to enable domain experts and novice users alike to discover complex correlations and to test hypotheses and differences between various populations in an entirely visual manner with just a few clicks; unfortunately, often ignoring even the most basic statistical rules. 
We recently performed a small user study and asked people to explore census data using such an interactive data exploration tool. 
Within minutes, all participants were able to extract multiple insights, such as ``\emph{people with a Ph.D. earn more than people with a lower educational degree}''. 
At the same time, almost none of the participants used a statistical method to test whether the difference the visually observed visually from the histogram is actually meaningful.
Further, most users including experts with statistical training, did not consider that this type of exploration, that consists of repeated attempts to find interesting facts, increases the chance to observe seemingly significant correlations by chance.

This problem is well known in the statistics community and referred to as the ``\emph{multiple testing problem}'' or ``\emph{multiple hypothesis error}'' and it denotes the fact the more tests an analysts performs, the higher is the chance that a discovery is observed by chance. 
Let us assume an analyst tests 100 potential correlations, 10 of them being true, and she wants to limit the chance of a false discovery to $5\%$ (i.e., the family-wise error rate should be $p=0.05$).
Assume further that our test has a statistical power (i.e, the likelihood to discover a real correlation) of 0.8; all very common values for a statistical testing.
With this setting, the user will find  $\approx 13$ correlations of which $5$ ($\approx 40$\%) are ``bogus''.  
The analyst should use a multiple hypothesis test correction method, such as the Bonferroni correction \cite{bonferroni1936teoria}.
However, Bonferroni correction significantly decreases the power of every test and with it the chance of finding a true insight. 
This is especially true in the case of interactive data exploration, where the number of tests is not known upfront and incremental versions of Bonferroni correction need to be applied which would even further decrease the power of the tests.

Another interesting question concerns what should be considered as a hypothesis test when users interactively explore data. 
For example, if a user sees a visualization, which shows no difference in salaries between men and women based on their education, but later on decides based on that insight to look at salary differences between married men and women. Should we still account for that? 
The answer in most cases will be ``yes'' as the analyst probably implicitly made a conclusion based on that visualization, which then in turn triggered her next exploration step.
However, if she considers this visualization just as a descriptive statistic of how the data looks like, and makes no inference based on it (i.e. 
it did not influence the decision process of what to look at next), then it should not be considered as a hypothesis. 
The difference is subtle and usually very hard to understand for non-expert users, while it might have a profound impact on the number of false discoveries a user makes. 

Finally, in the context of data exploration there has been recent work on automatically recommending visualization \cite{seedb,vizdeck,voyager} or correlations \cite{polygamy}. 
These systems yet again increase the chance of false discoveries since they automatically test all (or at least a large fraction) of possible combinations of features until something interesting shows up without considering the multiple hypothesis testing problem. 

In this paper, we make a first step towards integrating automatic multiple hypothesis testing control into an interactive data exploration tool.
We propose a potential user interface and a meaningful default hypothesis (i.e., the \emph{null hypothesis}), which allows us to achieve control of the ratio of false discoveries  for every user interaction. 
Specifically, we propose to consider every visualization as a hypothesis unless the user specifies otherwise. 
We further discuss control procedures based on the family-wise error and discuss why they are too pessimistic for interactive data exploration tools and why the more modern criteria of controlling the false discovery rate (FDR) is better suited for large scale data exploration. 
The challenge of FDR, however, is, that the standard techniques, such as the Benjamini-Hochberg procedure are not incremental and require to test all the hypotheses, before determining which hypotheses are rejected. This clearly constitutes a problem in the data exploration setting where users make discoveries incrementally. 
The recent {\em $\alpha$-investing} technique \cite{foster2008alpha} proposes an incremental procedure to control a variation of FDR, called marginal FDR (mFDR), 
which however relies on the user having a deep understanding of how valuable each individual test is supposed to be. 
Again a contradiction to data exploration, where the user only over time gains a feel about the importance of certain questions. 
We therefore propose new strategies based on the $\alpha$-investing procedure \cite{foster2008alpha}, which are particular tailored towards interactive data exploration tools. 
We implement these ideas in a system called \system{} and we show how this system can help experts and novice users alike to control false discoveries through extensive experiments on both real-world and synthetic data and workloads.

The main contributions can be summarized as follows:

\vspace{-2.0ex}
\begin{itemize*}
    \item We propose \system{}, a novel system which automatically tracks hypotheses during data exploration;
    \item We discuss several multiple hypothesis testing control methods and how well they work for data exploration;
    \item Based on the previous discussion, we develop  new $\alpha$-investing rules to control a variant of the false discovery rate (FDR), called marginal FDR (mFDR); 
    \item We evaluate our system using synthetic and real-world datasets and show that our methods indeed achieve control of the number of false discoveries when using an interactive data exploration system. 
\end{itemize*}
\vspace{-2.0ex}

The paper is structured as follows: 
in Section~\ref{sec:requirements} we discuss, by means of an example, why some visualizations should be considered hypothesis tests and what are the main challenges encountered when testing hypotheses for the IDE setting. 
In Section~\ref{sec:ui} we present \system{}'s user interface and discuss how to automatically track hypotheses and how to integrate the user feedback into tracking the hypothesis. 
In Section~\ref{sec:methods} we discuss multiple hypothesis testing techniques known in literature and show how well they fit in the IDE setting. 
In Section~\ref{sec:invest} we then propose new multiple hypothesis testing procedures for IDE based on the \ainv procedure. 
Afterwards, in Section~\ref{sec:eval} we present the result of our experimental evaluation using both real-world and synthetic data. 
Finally, in Section~\ref{sec:related} and \ref{sec:concl} we discuss related work and present our conclusions.

\begin{figure*}[ht]
\begin{center}
  \includegraphics[width=0.9\textwidth]{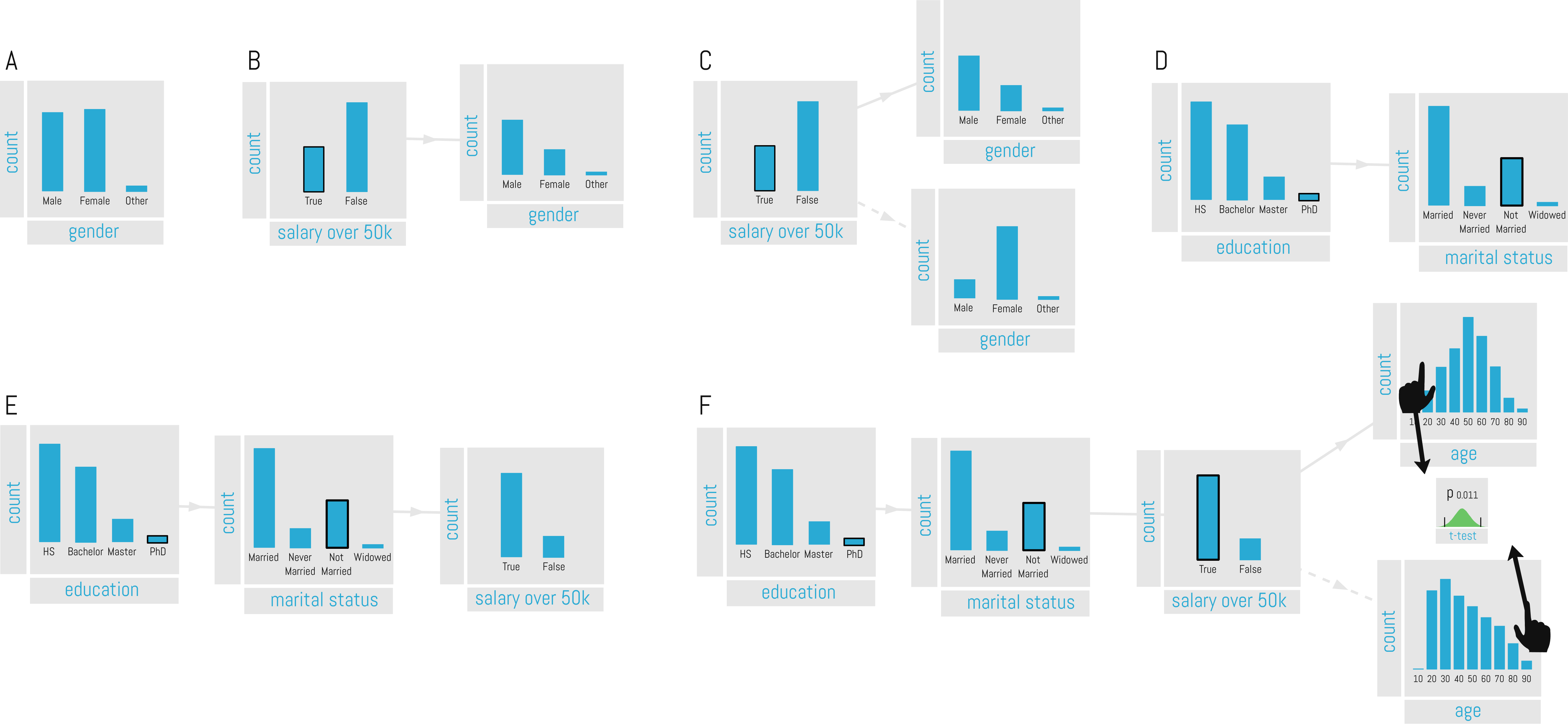}
\end{center}\vspace{-5.5ex}
\vspace{-2.5ex}
\caption{An example Interactive Data Exploration Session}
\vspace{-3.5ex}
\label{fig:sb}	
\end{figure*}

\section{A Motivational Example}
\label{sec:requirements}

To motivate the various aspects for multi-hypothesis control during data exploration we outline a potential scenario that is inspired by Vizdom \cite{vizdom}. Similar workflows however can be achieved with other systems like Tableau, imMens \cite{immens} or Dice \cite{dice}. 

Let us assume that Eve is a researcher at a non-profit organization and is working on a project relevant to a specific country. 
She just obtained a new dataset containing census information and is now interested in getting an overview of this data as well as extracting new insights. 
She starts off by considering  the ``\emph{gender}'' attribute and observes that the dataset contains the same number of records for men and women (Figure~\ref{fig:sb} A). 
She then moves to a second visualization, displaying the distribution of people who earn above or below \$50k a year. 
Eve links the two charts so that selections in the ``\emph{salary}'' visualization now filter the ``\emph{gender}'' visualization.
She notices that by selecting the salaries above \$50k, the distribution of ``\emph{gender}'' is skewed towards men, suggesting that men have higher salaries than women (B). 
After creating a third visualization for ``\emph{gender}'', selecting the records corresponding to records with salary lower than \$50k (dashed line indicates inversion of selection), she confirms her finding ``\emph{Women in this country are predominately earning less than \$50k}'' (C).
Eve now wants to understand what else influences a persons yearly salary and creates a chain of visualizations that selects people who have PhD degrees and are not married (D).
Extending this chain using the ``\emph{salary}'' attribute appears to suggest that this sub-population contains a lot of high-earners (E).
By selecting the high-earners and extending the chain with two ``\emph{age}'' visualizations allows her to compare the age distribution of unmarried PhDs earning more than \$50k to those making less than \$50k. 
In order to verify that the observed visual difference is actually  \emph{statistically significant} she performs a t-test by dragging the two charts close to each other (F).

While the example workflow contains only one hypothesis test \emph{explicitly} initiated  by the user, we argue that without accounting for other \emph{implicit hypothesis tests} there is a significantly increase of risk that the users may observe false discoveries during similar scenarios of data exploration. This opens up new important questions: why and when should visualizations be considered statistical hypothesis tests?  How should these tests be formulated?

\subsection{Hypothesis Testing} In this paper, we focus on the widely used frequentist inference approach and its \pval outcome. 
In order to determine whether there is a correlation between two observed phenomena formalized in a ``\emph{research hypothesis}'' $\mathcal{H}$ \ that is actually statistical relevant (i.e., not product of noise in the data) we analyze its corresponding ``\emph{null hypothesis}'' $H$  which refers to a general statement or default position according to which there is no relationship between two measured phenomena. Given this relationship between $\mathcal{H}$ and $H$, the research hypothesis $\mathcal{H}$ is also commonly referred as ``\emph{alternative hypothesis}''.

The \emph{testing procedure} will then determine whether to \emph{accept} (resp., \emph{reject}) a null hypothesis $H$ which in turn corresponds to rejecting (resp., accepting) the corresponding \emph{alternative hypothesis} (or \emph{research hypothesis}) $\mathcal{H}$. In order to do so the \pval of the null hypothesis $H$ is evaluated.
The \pval is used in the context of null hypothesis testing in order to quantify the idea of statistical significance of evidence and it denotes the probability of obtaining an outcome at least as extreme as the one that was actually observed in the data, under the assumption that $H$ is true. Depending on the context, the \pval of $H$ is evaluated using the appropriate statistical test (e.g., the \emph{t-test} or the $\Chi^2$\emph{-test}).  


If the \pval $p$ associated to the null hypothesis $H$ is less than or equal to the \emph{significance level} $\alpha$ chosen by the testing procedure (commonly $0.05$ or $0.01$), the test suggests that the observed data is inconsistent with the null hypothesis, so the null hypothesis must be rejected.
This procedure guarantees for a single test, that the probability of a ``\emph{false discovery}'' (also known as ``\emph{false positive}'' or ``\emph{Type I error}'') -- wrongly rejecting the null hypothesis of no effect -- is at most $\alpha$. 
This does not imply that the alternative hypothesis is true; it just states that the observed data has the likelihood of $p\leq \alpha$ under the assumption that the null hypothesis is true. 
The \emph{statistical power} or \emph{sensitivity} of a binary hypothesis test is the probability that the test correctly rejects the null hypothesis $H$ when the alternative hypothesis $\mathcal{H}$ is true.

While the frequentist approach to hypothesis test has been criticized \cite{jeffreys1998theory,neyman1948consistent} and there has been a lot of work in developing alternative approaches, such as Bayesian tests \cite{BH}, it is still widely used in practice and we consider it a good first choice to build a system which automatically controls the multiple hypotheses error as they have two advantages: 
(1) Novice users are more likely to have experience with standard hypothesis testing than the more demanding Bayesian testing paradigm. 
(2) The frequentist inference approach does not require to set a hard-to-determine {\em prior} as it is the case with Bayesian tests. 


\subsection{Visualizations as Hypotheses}
A visualization per-se shows a descriptive statistic (e.g., the count of women or the count of men) of the dataset and is not a hypothesis. 
It is reasonable to assume that in step~A of Figure~\ref{fig:sb} the user just looks at the gender distribution and simply acknowledges that the census surveys roughly the same amount of women and men.
However, it becomes an hypothesis test, if the user expected something else and draws a conclusion/inference based on the visualization. 
For example, if the user somehow assumed that there should be more men than women in the data and therefore considering the fact that there is an equal amount as an insight.
The notion of a visualization being considered as a hypothesis becomes even clearer in step~(B) and (C) of the example work-flow.
When looking at the visualization in (B) in isolation, it just depicts a descriptive statistic. 
Indeed, if the user would just take it as such and not make any inference about it and/or base further exploration on an insight extracted from this visualisation, then it would not be considered an hypothesis.
We argue however that the opposite is true more often than not. First, our analytical reasoning and sense-making process is inherently non-linear \cite{pirolli2005sensemaking,shrinivasan2008supporting}.
Our future actions are influenced by new knowledge we discovered in previous observations.
Second, while susceptible to certain types of biases \cite{attractionBias}, the human visual system is highly optimized at picking up differences in visual signals and at detecting patterns \cite{burgess1981efficiency}. 
An average user is very likely drawn to the changes between the gender distribution of step (A) and step (B) and might therefore infer that women earn less than men and potentially flag this as an interesting insight that deserves more investigation.
This is illustrated in step~(C) where the user now further drills down and visually compares the distribution of gender filtered by salary. 
We qualitatively confirmed this notion through a formative user study where we manually coded user-reported insights, following a think-aloud protocol similar to the one proposed in ~\cite{guo2016case}. In this study we observed that users tend to pick up on even slight differences in visualizations and regard them as insights and users predominantly base future exploration paths on previously inferred insights. 

We conclude two things: (1) most of the time users indeed treat visualizations as hypotheses, though there are exceptions, and (2) they often (wrongly) assume that what they see is statistical significant. 
The latter is particularly true if the users do not carefully check the axis on the actual count.  
For example, if a user starts to analyze the outliers of a billion record dataset and makes the conclusion that mainly uneducated whites are causing the outliers, the dataset she is referring to might be comparable small and the chance of randomness might be much higher. 
The same argument also holds against the critic, that with enough data observing differences by chance are much less likely, which is true. 
As part of visual data exploration tools, users often explore sub-populations, and while the original dataset might be large, the sub-population might be small. 
Thus, we argue that every visualization as part of a interactive data exploration tool should be treated as a hypothesis and that users should be informed about the significance of the insights they gain from the visualization. 
At the same time, a user should have the choice to declare a visualization as just descriptive. 


\subsection{Heuristics for Visualization Hypotheses}
A core question remains: what should the hypothesis for a visualization be. 
Ideally, users would tell the system every single time what they are thinking so that the hypothesis is adjusted based on their assumed insight(s) they gain from the visualization. 
However, this is disruptive to any interactive data exploration session. 
We rather argue that the system should use a good default hypothesis, the user can modify (or even delete) if she so desires. 
For the purpose of this work, we mainly focus on histograms as shown in Figure~\ref{fig:sb} and acknowledge that there exist many other visualizations, which we consider as future work. 
We derived the following heuristics from two separate user studies where we observed over 50 participants using a IDE tool to explore various datasets. 

\vspace{-1.75ex}
\begin{enumerate*}
    \item {\em Every visualization without any filter conditions is not a hypothesis (e.g., step~A in Figure~\ref{fig:sb}) unless the user makes it one. } This is reasonable, as users usually first gain a general high-level impression of the data. Furthermore, in order to make it an hypothesis, the user would need to provide some prior knowledge/expectation, for example as discussed before, that he expected more men than women in the dataset. 
    \item {\em Every visualization with a filter condition is a hypothesis with the null-hypothesis that the filter condition makes no difference compared to the distribution of the whole dataset}. For example, in step~B of Figure~\ref{fig:sb} the null hypothesis for the distribution of men vs. women given the high salary class of over $\$50k$ would be that there is no difference compared to the equal distribution of men vs. women over the entire dataset (the visualization in step~A). This is again a reasonable assumption as the distribution of an attribute given others is only interesting, if it shows some different effect compared to looking at the whole dataset. 
    \item {\em If two visualization with the same but some negated filter conditions are put next to each other, it is a test with the null-hypothesis that there is no difference between the two visualized distributions, which supersedes the previous hypothesis.} This is the case in step~C: given that the user looks explicitly at the distribution of males vs females given a salary over and under $\$50k$ is a strong hint from the user, that he wants to compare these two distributions. 
\end{enumerate*}
\vspace{-2.0ex}

As with every heuristic it is important to note, that the heuristic can be wrong. 
Therefore it is extremely important to allow the user to overwrite the default hypothesis as well as delete default hypothesis if one really just acted as a descriptive statistic or was just generated as part to a bigger hypothesis test. 
Furthermore, there exist of course other potential null-hypothesis.
For example, in our workflow we assume by default that the user aims to compare distributions, which requires a  $\chi^2$-test.
However, maybe in some scenarios comparing the means (i.e., a t-test) might be more appropriate as the default test. Yet, studying in detail what a good default null-hypothesis is dependent on the data properties and domain, is beyond the scope of this paper. 

\subsection{Heuristics Applied to the Example}
For our example in Figure~\ref{fig:sb} the resulting hypothesis could be as follows:
Step~A is not an hypothesis based on rule 1 as it just visualizes the distribution of a single attribute over the whole dataset. 
Step~B is the hypothesis $m_1$ if the distribution of gender is different given a salary over $\$50k$. 
Step~C supersedes the previous hypothesis and replaces it with an hypothesis $m_1'$ if the gender distribution between a salary over and under $\$50k$ is different, which is a sightly different question. 
Step~D creates a hypothesis $m_2$ if the marital status for people with PhDs is different compared to the entire dataset, whereas step-E generates a hypothesis $m_3$ if there is a different salary distribution given not married people with a PhD. 
By studying the age distribution in step~F the system first generated a default hypothesis $m_4$  that the distribution of the ages is different given a PhD and being not married for different salary classes. 
However, the user overwrites immediately the default hypothesis with an hypothesis $m_4'$  about the average age. 
Furthermore, as the previous visualizations in step~D and E might just have been stepping stones towards creating $m_4‘$ the user might or might not delete hypothesis $m_2$ and $m_3$. 
However, if the insights our user gained from viewing the marital status, etc., influenced her to look at the age distribution, she might want to keep them as hypothesis. 

Clearly this is only a very small example, but it already demonstrates the general issues. 
Not every insight the user gains (e.g., the insight that women earn less) is explicitly expressed as a test. 
At the same time, as more the user ``surfs'' around the higher the chance that she finds something which looks interesting, but just appears because of chance. 
In the example above, by the time the user actually performs its first test (step~F), she implicitly already tested at least one other hypothesis and potentially even four others. 
Assuming a targeted \pval of $\alpha = 0.05$, the chance of a false discovery therefore increased to $1 - (1 - \alpha)^2=0.098$ for two hypothesis and up to $1 - (1 - \alpha)^4=0.185$ for four hypothesis. 
While the question of what should count as an hypothesis is highly dependent on the user and can never be fully controlled by any system, we can however, enable the system to make good suggestions and help users to track the risk of making false discoveries by chance. 
Furthermore, this short workflow also demonstrates that hypotheses are built by adding but also by removing attributes. 
As we will discuss later, there exist no good method so far to control the risk of making false discoveries for incremental sessions like the ones created by interactive data exploration systems. 
We therefore develop new methods especially for interactive data exploration in Section~\ref{sec:invest}.

Finally, it should be noted, that the same problems also exist with exploratory analysis using SQL or other tools. 
However, we argue that the situation is becoming worse by the up-rise of visual exploration tools, like Tableau, which are often used by novice users, who not necessarily reflect enough on their exploration path after they found something interesting.




\section{The {\Large \system{}} User Interface}
\label{sec:ui}

As argued in the previous section, user feedback is essential in determining, tracking and controlling the right hypothesis during the data exploration process.
With \system~we created a system that applies our heuristic automatically to all visualizations. We designed \system~'s user interface with a few goals in mind.

First, the user should be able to see the hypotheses the system assumed so far, their \pvals, effect sizes and if they are considered significant and should be able to change, add or delete hypotheses at any given stage of the exploration. 

Second, hypotheses rejection decisions should never change based on future user actions unless the user explicitly asks for it. We therefore require an incremental procedure to control the multiple hypothesis risk that does not change its rejection decisions even if more hypothesis tests are executed.
For example, the system should not state that their is a significant age difference for not married highly educated people, and then later on revoke its assessment just because the user did more tests. 
More formally, if the system determined which hypotheses $m_1 ...  m_n$ are significant (i.e., it rejects the null) or not and the user changes the last hypothesis or adds an hypothesis $m_{n+1}$, which should be the most common cases, the significance of hypotheses $m_1..m_{n}$ should not change. 
However, if the user might change, delete, or add hypothesis $k \in {1,..,n}$, depending on the used procedure we might allow that the significance of hypotheses $m_{k+1}$ to $m_n$ might have to change as well.

Third, individual hypothesis descriptions should be augmented with information about how much data $n^{H1}$ the user has to add, under the assumption that the new data will follow the current observed distribution of the data, to make an hypothesis significant. 
While sounding counter-intuitive, as one might (wrongly) imply, it is possible to make any hypothesis true by adding more data, calculating this value is in some fields already common practice. 
For example, in genetics scientist often search (automatically) for correlations between genes and high-level effects (like cancer). 
If such a correlation is found, often because of the multiple hypothesis error the chance of a true discovery is tiny (i.e., the \pval is too high). 
In that case the scientist works backwards and estimates how much more genes she has to to sequence in order to make the hypothesis relevant, expecting that the new data (e.g., gene sequences) follow the same distribution of the data the scientist already has.
However, if the effect was just produced by chance, the new data will be more similar to the distribution of the null-hypothesis and the null will not be rejected.  
The required value is generally easy to calculate or approximate,  and are highly valuable for the end-user. 
A small value for $n^{H1}$ in relation to the number of totally tested hypotheses might be an indication that the power (i.e., the chance to accept a true alternative hypothesis) of the test was not sufficiently large. 

And finally, users should be able to bookmark important hypotheses. 
Our system uses default hypothesis throughout the exploration and the user might find it too cumbersome to correct everyone for his real intentions, there might be more hypotheses generated than the user intended to test. 
Even if all hypotheses are what the user was considering, some of them might be more important to her than others; the hypotheses the user would like to include in a presentation or show to her boss. 
A key key question becomes, what is the expected number of false discoveries among those important discoveries?

\begin{figure}
\begin{center}
\hspace*{-7.5ex}
\includegraphics[width=0.53\textwidth]{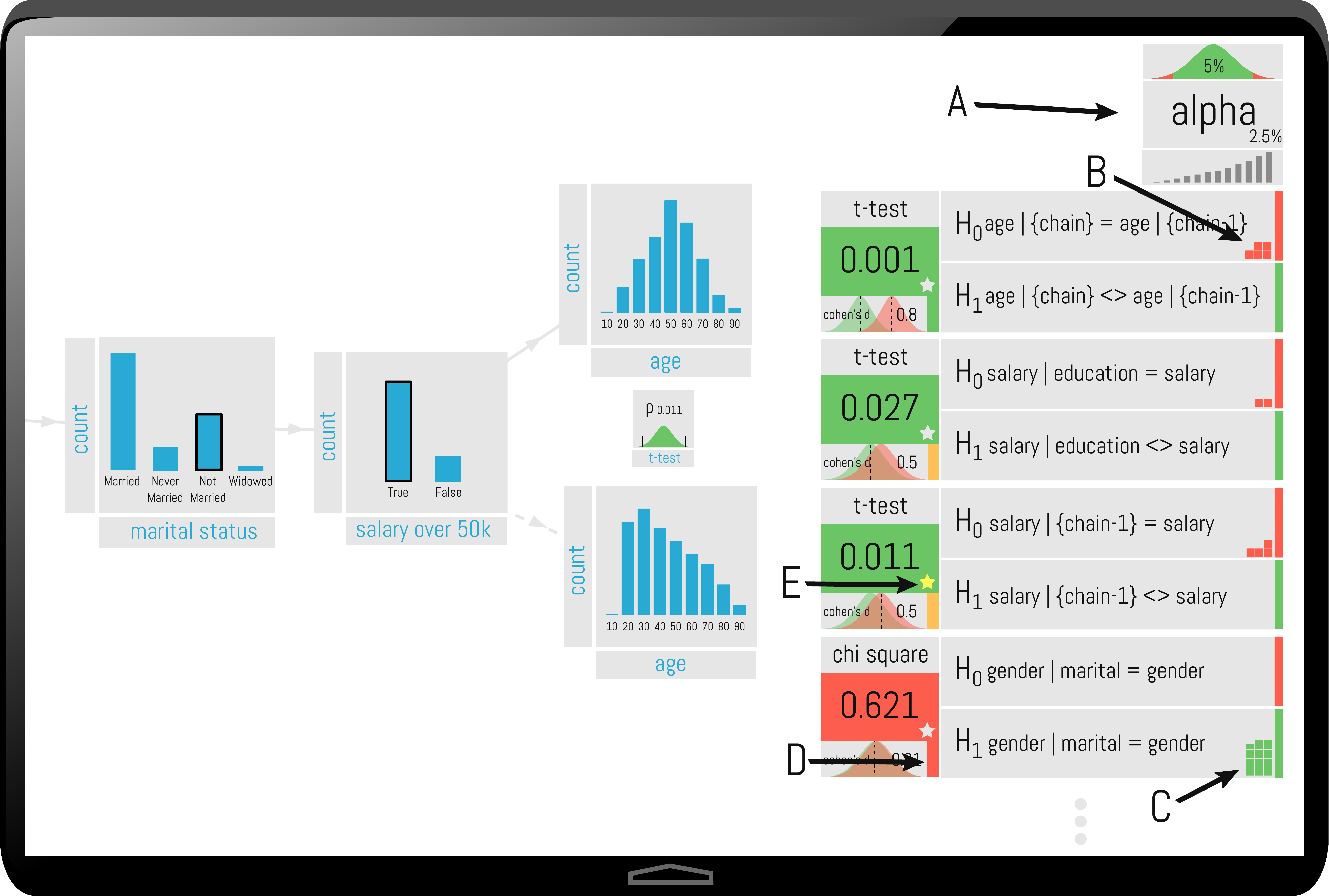}
\end{center}
\vspace{-3.5ex}
\caption{The \system{} User Interface}
\vspace{-5.5ex}
\label{fig:riskcontroller}	
\end{figure}

Figure \ref{fig:riskcontroller}	shows the current interface design of \system{} with a risk controller, which incorporates the above ideas, running on a tablet. 
The user interface features an unbounded 2D canvas where chains of visualizations (such as the one shown in Figure \ref{fig:sb}) can be laid out in a free form fashion. 
A ``risk-gauge'' on the right-hand side of the display (Figure \ref{fig:riskcontroller} (A)) serves two purposes: 
it gives users a summary of the underlying procedure (e.g., the budget for the false discovery rate set to 5\% with current remaining wealth of 2.5\%; 
both explained in the next two sections) and it provides access to a scrollable list of all the hypothesis tests (implicit and explicit) that have been execute so far. 
Each list entry displays details about one test and its results.
Textual labels describe the null- and alternative-hypothesis and color coded \pvals  indicate if the null-hypothesis was rejected or accepted (green for rejected, red for accepted).  
Furthermore, it visualizes the distribution of null-hypothesis and alternative hypothesis and shows its difference, included an indication of its color coded effect size (D). 
Tap gestures on a specific item allow users to change things like the default hypothesis or the type of test.  
Additionally other information such as an estimation of the size of an additional data  $n^{H1}$ that could make the observation significant can be displayed in each item. 
In the example this information is encoded through a set of small squares (B, C) where each square indicates the amount of data that is in the corresponding distribution. In (B) the five red squares tells us that we need 5x the amount of data from the null-distribution to flip this test form rejected to accepted or conversely in (C) 11.5x the amount of data from the alternative-distribution to rejected this hypothesis. 
Finally, we allow to mark important hypotheses by tapping the ``star'' icons (E). 






\section{Background on Multiple Hypothesis Error}
\label{sec:methods}

The previous section described how we convey the multiple hypothesis error to the user and ask for user feedback to derive the right hypothesis. 
In this section we describe different alternatives to calculate the potential false discovery error and discuss they appropriateness for the IDE setting. 
The notation used in the rest of the paper is summarized in Appendix \ref{appendix:symbols}.

We consider a 
setting, in which we evaluate the statistical relevance of hypotheses from a set $\mathcal{H}={\mathcal{H}_1,\mathcal{H}_2,\ldots,\mathcal{H}_m}$, created incrementally by an IDE system in a streaming fashion.
In order to verify whether any such hypothesis $\mathcal{H}_j$ is in fact statistically relevant we consider its corresponding null hypothesis $H_j$. 
Using the appropriate statistical test (e.g., the \emph{t-test} or the $\Chi^2$\emph{-test}) the \pval of $H_j$ evaluated and based on it the testing procedure determine whether to \emph{accept} (resp., \emph{reject}) a null hypothesis $H_j$ which in turn corresponds to rejecting (resp., accepting) the corresponding \emph{alternative hypothesis} (or \emph{research hypothesis}) $\mathcal{H_j}$.
The hypothesis according to which all null hypotheses are true is referred as the ``\emph{complete}'' or ``\emph{global}'' null hypothesis.

The set of null hypotheses rejected by a statistical test are called ``\emph{discoveries}''and are denoted as $R$. 
Among these we distinguish the set of \emph{true discoveries} $S$, and the set of \emph{false discoveries} or \emph{false positives} $V$; i.e., $|V|+|S|= |R|$ 
False discoveries are commonly referred also as Type 1 errors. 
Null hypotheses in $S$ are \emph{false null hypotheses}, while null hypotheses in $V$ are \emph{true null hypotheses}. 

\subsection{Hold-Out Dataset}
A possible method to deal with  the multiple hypothesis error is to split the dataset $D$ into a exploration $D_1$ and a validation $D_2$ dataset \cite{MEE3:MEE31}. 
$D_1$ is then used for the data exploration process, whereas the validation dataset is used to re-test all hypotheses in order to \emph{validate} the results of the first phase. 
In the following we will provide some examples which will clarify how, albeit useful,  a hold-out dataset does not solve the multiple hypothesis testing problem.

Let us consider a null hypothesis $H$, and let $p_D$ denote its associated \pval when $H$ is evaluated with respect of the entire dataset $D$. Lets assume we perform a test with significance-level $\alpha$. In this case the probability of wrongly rejecting $H$ is at most $\alpha$
Suppose now that we randomly split the dataset into two datasets $D_1$ and $D_2$. For the same null hypothesis $H$ we evaluate the \pvals $p_{D_1}$ and $p_{D_2}$ each obtained by evaluating $H$ on $D_1$ or $D_2$ respectively. We then run a a test with significance-level $\alpha$ (like the one discussed above) for each of the datasets. We then decide to reject $H$ if it has been rejected by \emph{both} the testing procedures operating on the datasets $D_1$ or $D_2$. 
If both procedures operating on $D_1$ and $D_2$ have significance-level $\alpha$, then the probability that the overall procedure ends up rejecting $H$ is at most $\alpha^2$.


For the common value of $\alpha=0.05$, the chance of a Type I error is thus reduced to  $0.0025$, which is good news. 
Rather than fully handling the multiple hypothesis problem, what we have achieved trough this procedure is however just the lowering of the threshold for rejecting the null hypothesis (i.e., the significance level of the test).

This fact appears clearly in the following scenario.
Suppose that the user wants to evaluate multiple hypotheses (e.g., 25) rather than just one. Assuming that these hypotheses, and their \pvals are independent, the probability of observing at least one erroneous rejection using the test technique based on the use of the holdout dataset would be:
$p_f = 1 - (1-p_D)^{25} \approx 0.06$,
which is higher than the desired $\alpha$ significance level. 

Albeit the lowering of the achieved reduction of the significance level is indeed useful for reducing the chance of Type I errors, it comes at the cost of a significant reduction of the power of the testing procedure.

Let us consider the following example scenario in which we aim to compare the means $M_1$ and $M_2$ of two samples one drawn from a population with expected value $\mu_1=0$ and the other from a population with $\mu_2=1$, both having a standard deviation of $\sigma=4$. 
In order to determinate weather the observed difference between $M_1$ an $M_2$ is actually statistically significant, we test the null-hypothesis ``\emph{there is no  significant difference between $\mu_1$ and $\mu_2$}'' using the \emph{one-sided t-test} and a sample composed by 500 records from each population. Given the properties of the t-test (see~\cite{fisher1935}), the statistical power of our test would be $0.99$, and the probability of erroneously accepting the null hypothesis would be at most $0.01$.

Suppose now that we divide the dataset into a dataset for exploration and one for validation each composed by 250 records. The statistical power for each of the individual t-test executed on the two dataset is now lowered to $0.87$, due to the reduction of the data being used.
Further, recall that the procedure based on the holdout set rejects a null hypothesis only if said hypothesis is rejected by both sub-tests. This implies that the actual \emph{overall} power of the testing procedure is $0.87 \cdot 0.87 \approx 0.76$, which is significantly lower than the $0.99$ achieved by the test which uses the entire data. 

In general, approaches based on hold-out datasets are considered inferior compared to testing over the entire dataset. 
In some scenarios, like building machine learning models, hold-out datasets might even be the only possibility to test a model or tune parameters. 
In those cases, a hold-out approach (like k-fold cross-validation) should be considered as test and should be controlled for the multiple hypothesis error as recent work suggests \cite{Demsar2006,Kohavi1995,Refaeilzadeh2009}.

It is however important to remark that in our work we aim to predict guarantees on the statistical significance of the statistical predictors which are instead not achievable using prediction-driven approaches such as cross-validation.

\subsection{Family-Wise Error Rate (FWER)}
Traditionally, frequentist methods for multiple comparisons testing focus on correcting for modest numbers of comparisons.
A natural generalization of the significance level to multiple hypothesis testing is the {\it Family Wise Error Rate}, which is the probability of incurring at least one Type I error in any of the individual tests. 
The FWER is the probability of making at least one type I error in the family:

\vspace{-4.5ex}\begin{equation}\scriptsize\scriptsize
FWER =\Pr(V\geq 1) = 1-\Pr(V=0)
\end{equation}\vspace{-3.5ex}

By assuring that $FWER\leq \alpha$, that is the FWER is \emph{controlled at level} $\alpha$, we have that the probability of even one Type I error in evaluating a family of hypotheses is at most $\alpha$.

We say that a procedure controls the FWER \emph{in the weak sense}, if the FWER control at level $\alpha$ is guaranteed only when \emph{all} null hypotheses are true (i.e. when the complete null hypothesis is true).
We say that a procedure controls the FWER \emph{in the strong sense}, if the FWER control at level $\alpha$ is guaranteed for any configuration of true and non-true null hypotheses (including the global null hypothesis).

{\bf Bonferroni Correction: }
The Bonferroni correction is the simplest statistical procedure for multiple hypothesis testing~\cite{bonferroni1936teoria}. Let $\alpha$ be the critical threshold for the test. 
The value of $\alpha$ is usually selected at $0.01$ or $0.05$.

Let $p_i$ the \pval statistic associated with the null hypothesis $H_i$. When testing $m$ distinct null hypotheses using the Bonferroni correction, a null hypothesis $H_i$ is rejected if $p_i\leq \alpha/m$. The Bonferroni procedure thus achieves control of the FWER at level $\alpha$.



Unfortunately, the Bonferroni correction can not be applied in our setting as it requires knowledge of the total number of hypotheses being considered. 
An alternative approach is to use a variation of the Bonferroni correction, according to which the $j$-th null hypothesis $H_j$ is rejected if $p_j\leq \alpha \cdot 2^{-j}$. It is possible to show that this procedure indeed controls FWER at level $\alpha$ as $j\rightarrow \infty$ and does not need explicit knowledge of $m$. However the acceptance threshold decreases exponentially with respect to the number of hypotheses, thus resulting in a high number of false negatives. 

The main common issue with all FWER techniques is that the power of the test significantly decreases as $m$ increases due to the corresponding decrease in the acceptance threshold ($\alpha/m$ in the original Bonferroni or $\alpha/2^i$ in the sequential variant).
While some alternative testing procedures such as those of V\v{i}d\'{a}k~\cite{vsidak1967rectangular}, Holm~\cite{holm1979simple}, Hochberg~\cite{hochberg1988sharper}, and Simes~\cite{simes1986improved} offer more power while controlling FWER, the achieved improvements are generally minor. A review of several of these techniques is provided by Shaffer in~\cite{shaffer1995multiple}.

\subsection{False Discovery Rate (FDR)}
In~\cite{BenjaminiH95}  Benjamini and Hochberg proposed the notion  of {\it False Discovery Rate (FDR)} as a less conservative approach to control errors in multiple tests which achieve a substantial increase in the power of the testing procedure. 

FDR-controlling procedures are designed to control the expected ratio $Q=V/R$ of false discoveries among all discoveries returned by a procedure. In particular, the FDR of a statistical procedure is defined as:

\vspace{-4.5ex}\begin{equation}\label{eq:fdr1}\scriptsize
FDR = E\left[Q\right] = E\left[\frac{V}{R}|R>0\right]P(R>0).
\end{equation}\vspace{-2.5ex}

However, if we define FDR to be zero when $R=0$, we can simplify \ref{eq:fdr1} to:

\vspace{-4.5ex}\begin{equation}\scriptsize\label{eq:fdr}
FDR=E\left[\frac{V}{R}\right]
\end{equation}\vspace{-2.5ex}




We say that a testing procedure controls FDR at level $\alpha$ if we have $FDR \leq \alpha$.
Designing a statistical test that controls for FDR is not simple, as the FDR is a function of two random variables that depend both on the set of null hypotheses and the set of alternative hypotheses.
The standard technique to control the FDR is  the \emph{Benjamini-Hochberg procedure}(BH), which operates as follows:
let $p_1\leq p_2\leq \ldots\leq p_m$ be the sorted order of the 
the \pvals for the $m$ tested null hypotheses. 
To control FDR at level $\alpha$ (for independent null $p$-values) determine the maximum $k$ for which $p_k\leq \frac{k}{m}\cdot\alpha$, and reject 
the null hypotheses corresponding to the \pvals $p_1,p_2,\ldots,p_k$.

Interestingly, under the complete null hypothesis, controlling the FDR at level $\alpha$ guarantees also  ``\emph{weak control}'' over the FWER $FWER=P\left(V\geq 1\right)=E\left({\frac {V}{R}}\right)=\mathrm {FDR} \leq \alpha$. This follows from the fact that the event of rejecting at least one true null hypothesis $V\geq 1$ is exactly the event $V/R=1$, and the event $V=0$ is exactly the event $V/R=0$ (recall $V/R=0$ when $V=R=0$). 
This makes the FDR relatively easy to explain to the user as under complete random data, the chance of one or more false discoveries is at most $\alpha$ as in FWER. 
However, FDR does not however ensure control of the FWER if there are some true discoveries to be made (i.e., it does not ensure ``\emph{strong control}'' of the FWER). 

Because of its increased power, FDR appears to be a better candidate than FWER in the context interactive data exploration, where usually a larger number of hypotheses are to be considered. 
Unfortunately, both  the original \emph{Benjamini-Hochberg procedure} and its variation for dealing with dependent hypotheses~\cite{BY01} are not incremental as they require knowledge of the total number of hypotheses being tested (similar to what was discussed for Bonferroni) and of the sorted list of all the \pvals corresponding to each null hypothesis being evaluated. 

An adaptation of the FDR technique to a setting for which an unspecified number of null hypotheses are observed incrementally was recently discussed in~\cite{GSell}. The main idea behind the Sequential FDR procedure is to convert the arbitrary sequence of \pvals corresponding to the null hypotheses observed on the stream of hypotheses into an ordered sequence akin to the one generated by the classical Benjamini-Hochberg procedure. The natural application for this technique is the progressive refinement of a model by considering additional features. That is, it starts constructing a model for the data with something known and general. The user then proceeds to refine the model by determining the most significant features.

One drawback of the Sequential FDR method, is given by the fact that the order according to which the hypotheses are observed on the stream heavily influences the outcome of the procedure. For example, if an hypothesis with high \pval is observed among the first in the stream, this will harm the ability of the procedure of rejecting following null hypotheses, even if they have low \pval (see discussion in~\cite{GSell}). This aspect makes Sequential FDR not applicable for data exploration system for which the user is likely to explore different ``\emph{avenues}'' of discovery rather than focusing on the specialization of a model.

\subsection{Other Approaches}

Although for most practical applications, FDR controlling procedures constitute the \emph{de facto standard} for multiple hypothesis testing~\cite{efron2016computer}, many other techniques have been presented in the literature. 
Among them, Bayesian techniques are particularly noteworthy. In~\cite{BH}, alternative solutions to the multple hypothesis problem combining decision theory with Bayesian FDR are discussed.
However, as often the case with Bayesian approaches, the computational cost for these procedures when applied to large datasets are significant, and the results are highly dependent on the prior model assumptions.

Another approach is correcting for the multiplicity through simulations (e.g., the \emph{permutation test}~\cite{schemper1984survey}) that experimentally evaluate the probability of an observation in the null distribution. This approach is also not practical in large datasets because of the large number of different possible observations and the need to evaluate very small $p$-values of each of these distributions~\cite{Jor11}. 

In this paper, we elect to use a family of multiple hypothesis testing procedures know as \ainv introduced in~\cite{foster2008alpha} and then generalized in~\cite{aharoni2014generalized}. 
These procedures are especially interesting for the incremental and interactive nature of interactive data exploration. 
The details of \ainv and its application to our setting is extensively discussed in the next section.

\section{Interactive Control using {\Large {$\alpha$}}-Investing}
\label{sec:invest}

One drawback of the Sequential FDR procedure~\cite{GSell} as well as adaptations of FWER controlling techniques to the streaming setting is given by the fact that decisions regarding the rejection or acceptance of previously considered null hypotheses could potentially be overturned in latter stages due to new hypotheses being considered. Although statistically sound, this fact could appear extremely counter intuitive and confusing to the user. The only way to adopt the Sequential FDR procedure to data exploration would be to batch all the hypotheses and only present the final decisions afterwards. In that sense Sequential FDR is incremental but non-interactive in data exploration.

In order to have both incremental and interactive multiple hypothesis error control, we consider a different approach for multiple hypothesis testing based on the ``\ainv'' testing procedure introduced originally introduced by Foster and Stine in~\cite{foster2008alpha}. Similarly to \sfdr, this procedure does not require explicit knowledge of the total number of hypotheses being tested and can therefore be applied in the hypothesis streaming setting. \ainv presents however several crucial differences with respect to both traditional and sequential FDR control procedures. 

In the following, we first introduce the general outline of the procedure as presented in \cite{foster2008alpha} and then discuss several investing strategies (called policies) that we have developed for interactive data exploration.

\subsection{Outline of the Procedure}\label{sec:aoutline}

For \ainv, the quantity being controlled is not the classic FDR but rather an alternative quantity called ``\emph{marginal FDR (mFDR)}'':

\vspace{-2.5ex}\begin{equation}\scriptsize
mFDR_{\eta}(j) = \frac{\Ex{V(j)}}{\Ex{R(j)}+\eta}
\end{equation}\vspace{-2.5ex}

\noindent where $j$ denotes the total number of tests which have been executed, while $V(j)$ (resp., $R(j)$) denote the number of false (resp., total) discoveries obtained using the \ainv procedure. 

In particular, we say that  a testing procedure controls $mFDR_{\eta}$ at level $\alpha$ if $mFDR_{\eta}(j)\leq \alpha$.
The parameter $\eta$ is introduced in order to weight the impact of cases for which the number of discoveries is limited. Common choices for $\eta$ are $1,(1-\alpha)$, whereas the procedure appears to lose in power for values of $\eta$ close to 0~\cite{foster2008alpha}.

Under the complete null hypothesis we have $V(j)=R(j)$ hence $mFDR_{\eta}(j)\leq \alpha$  implies that $\Ex{V(j)}\leq \alpha \eta / \left(1-\alpha\right)$. If we chose $\eta=1-\alpha$ then $\Ex{V(j)}\leq \alpha$, and we can thus conclude that control of the $mFDR_{1-\alpha}$ at level $\alpha$ implies weak control fo the FWER at level $\alpha$~\cite{foster2008alpha}.
We refer the reader to the original paper of Foster and Stine~\cite{foster2008alpha} for an extensive discussion on the relationship between $mFDR$ and the classic FDR. A generalization of the \ainv procedure was later introduced in~\cite{aharoni2014generalized}.  
The \ainv procedure does not in general require any assumption regarding the independence of the hypotheses being tested, although opportune corrections are necessary in order to deal with possible dependencies. 
In our analysis, we however assume that all the  hypotheses and the corresponding \pvals are indeed independent.

Intuitively the $\alpha$-investing procedure works as follows:
With every test $j$ the users sets an $\alpha_j$-value, which has to be below the current wealth, which is in the beginning usually $\alpha \cdot (1-\alpha)$ before he performs the test. If the null-hypothesis is accepted ($p_j > \alpha_j$) the invested alpha value is lost. 
To some degree this is similar to the Bonferroni-correction as one could consider the $\alpha_j$ value everybody is compared to as  $\alpha/m$.
So whenever a test is performed, the wealth decreases by $\alpha/m$ until the wealth is 0 and the user has to stop exploring. 
However, in contrast to the Bonferroni-correction, with $\alpha$-investing the user can regain wealth through a rejected null-hypothesis, which makes the procedure truly incremental as it does no longer depend on the number of anticipated hypotheses $m$ and also more powerful. 

More formally, we denote as $W(0)$ the initial $\alpha$-wealth assigned to the testing procedure. If the goal of the testing procedure is to control $mFDR_{\eta}$ at level $\alpha$, then we shall set $W(0)=\alpha \cdot \eta$.
Here, $\eta$ is commonly set to $(1-\alpha)$. 
We denote as $W(j)$ the amount of ``\emph{available }$\alpha$-\emph{wealth}'' after $j$ tests have being executed.



Each time a null hypothesis $H_j$ is being tested, it is assigned a budget $\alpha_j>0$. Let $p_j$ denote the \pval associated with the null hypothesis $H_j$.
This hypothesis is rejected if $p_j\leq \alpha_j$. 
If $H_j$ is rejected than the testing procedure obtains a ``\emph{return}'' on its investment $\omega \leq \alpha$. 
Instead, if the null hypothesis $H_j$ is accepted, $\alpha_j/(1-\alpha_j)$ alpha wealth is deducted from the available $\alpha$-wealth:

\vspace{-2.5ex}\begin{equation}\scriptsize\scriptsize\label{eq:alpharul}
\scriptsize
W(t) - W(t-1) = \begin{cases}
\omega &\text{ if }p_j\leq \alpha_j,\\
-\frac{\alpha_j}{1-\alpha_j} &\text{ if }p_j> \alpha_j
\end{cases}
\end{equation}\vspace{-2.5ex}

\noindent The testing procedure halts when the available $\alpha$-wealth reaches $0$. 
At that point in time, the user should stop exploring to guarantee that $mFDR \le\alpha$.
Obviously again something, which is not desirable as it is hard to convey to any user, that he has to stop exploring. 
We will discuss this problem and potential solutions in Section~\ref{sec:fdrcontrol:nowealth}.

The budget $\alpha_j$ which can be assigned to test must be such that regardless of the outcome of the test, the available $\alpha$-wealth available after the test is not negative $W(j)\geq 0$, hence $\alpha_j\leq W(j-1)/\left(1-W(j-1)\right)$. Further we impose that $\alpha_j < 1$. While this constraint was not explicated in~\cite{foster2008alpha}, it is indeed necessary for the correct functioning of the procedure. Setting $\alpha_j = 1$ would lead to the potential deduction of an infinite amount of $\alpha$-wealth, violating the non negativity of $W(j)$. Setting $\alpha_j > 1$ would instead lead to having a positive increase of the available $\alpha$-wealth \emph{regardless} of the outcome of the test. In our analysis we will however assume that all the hypotheses being considered are indeed independent and their associated \pvals are independent as well.

We refer as ``$\alpha$-\emph{investing rule}'' to the policy according to which available budget has to be assigned to the hypotheses that needs to be tested.
Furthermore, in~\cite{foster2008alpha} it was shown that any $\alpha$-investing policy for which $W(0)=\eta \cdot \alpha$, $\omega = \alpha$, and which obeys the rule in (\ref{eq:alpharul}), controls the $mFDR$ at level $\alpha$, for $\alpha,\eta\in[0,1]$.

The freedom of assigning to each hypothesis a specific level of confidence independent of the order, and the possibility of ``\emph{re-investing}'' the wealth obtained by previous rejection constitute great advantages with respect to the  Sequential FDR procedure.


\subsection{$\alpha$-Investing for Data Exploration}

While it is relatively straightforward to devise investing rules, it is difficult a priori to determinate the ``\emph{best way to invest}'' the available $alpha$-wealth.  
If $\alpha_j$ is picked too small, the statistical power of every test is reduced and the chance is even higher too loose the invested wealth given a true alternative hypothesis. 
If $\alpha_j$ is too large, the entire $\alpha$ wealth might be quickly exhausted and the user (in theory) has to stop exploring or re-evaluate all his test (see also Section~\ref{sec:fdrcontrol:nowealth}). 
A policy is most likely to be successful if it can exploit some knowledge of the testing setting. 

Another complication is the construction of tests for which one can obtain the needed \pvals. To show that a testing procedure controls $mFDR$, we require that conditionally on the prior j - 1 outcomes (denoted as $R_i$), the level of the test of $H_j$ must not exceed $\alpha_j$: 

\vspace{-4.5ex}\begin{equation}\scriptsize
P(R_j = 1 | R_{j−1}, R_{j−2}, . . . ,R_1) \leq \alpha_j.
\end{equation}\vspace{-2.5ex}

This does not however constitute a problem in our setting as we are assuming all hypotheses and their \pval to be independent.

While \cite{foster2008alpha} proposed various investing rules, most of the proposed procedures might test a hypothesis again and overturn an initial rejection of a null-hypothesis. 
Therefore, in the remainder of this section we propose different \ainv policies particular for Interactive Data Exploration, which correspond to different exploration strategies and at exploiting different possible properties of the data.
However it should be noted, that our first procedure, $\beta$-farsighted, is a generalization of the ``\emph{Best-foot-forward policy}'' in \cite{foster2008alpha}.

For this paper, we consider a setting for which we observe a (potentially infinite) stream of null hypotheses for which at each of the discretized time steps a new null hypothesis is observed on the stream. We denote as $H_j$ the hypothesis being considered at the $j$-th step. We further assume that said hypotheses are independent. 

All our policies assign to each hypothesis a strictly positive budget $\alpha_j>0$ as long as any $\alpha$-wealth is available. If $p_j\leq \alpha_j$, the null hypothesis $H_j$ is rejected (i.e., it is considered a discovery). Vice versa, if $p_j> \alpha_j$ is \emph{accepted}. The current $\alpha$-wealth $W(j)$ is then updated according to the rule in (\ref{eq:alpharul}) and because of it controls $mFDR$ at level $\alpha$ as shown in \cite{foster2008alpha}. 

\subsection{$\beta$-Farsighted Investing Rule}



Like with real investment, the question is if one should invest short or long-term. 
With $\beta$-farsighted we created a policy, which tries to preserve wealth over long exploration sessions. 
Given $\beta\in [0,1)$, we say that a policy is $\beta$-farsighted if it ensures that regardless of the outcome of the $j$-th test at least a fraction $\beta$ of the current $\alpha$-wealth $W(j-1)$ is preserved for future tests, that is for $j=1,2,\ldots$:

\vspace{-2.5ex}\begin{equation}
\scriptsize
\begin{split}
W(j) &\geq \beta W(j-1),\\
W(j)- W(j-1) &\geq (\beta-1) W(j-1)
\end{split}
\end{equation}\vspace{-2.5ex}

We therefore define the $\beta$-farsighted procedure to control  $mFDR_{\eta}$ at level $\alpha$ in the procedure for Investing Rule \ref{alg:beta}.

\begin{algorithm}
\scriptsize
   \caption{$\beta$-farsighted}
   \label{alg:beta}
    \begin{algorithmic}[1]
        \State $W(0) = \eta \alpha$
        \For{$j = 1,2,...$ } 
            \State $\alpha_j = \min\left(\alpha, \frac{W(j-1)\left(1-\beta\right)}{1 + W(j-1)\left(1-\beta\right)}\right)$
            \If{$p(H_j) < \alpha_j$}
                \State $W(j)= W(j-1)+\omega$
             \Else
                \State $W(j)= W(j-1) - \frac{\alpha_j}{1 - \alpha_j} = \beta W(j-1)$
            \EndIf
        \EndFor

\end{algorithmic}
\end{algorithm}

 Different choices for the parameter $\beta \in [0,1)$ characterize how conservative the investing policy is. If there is high confidence on the first observed hypotheses being true discoveries, small values of beta (i.e., 0.25) would be more effective. Vice versa, high values of $\beta$ (i.e. 0.9) ensure that even if the first hypotheses are true null, a large part of the $\alpha$-wealth is preserved.
 
We say that an $\alpha$ investing policy is ``\emph{thrifty}'' if it never fully commits its available $\alpha$-wealth. The described $\beta$-farsighted is indeed thrifty. While the  procedure will never halt due to the available $\alpha$-wealth reaching zero, after a long series of acceptance of null hypotheses the available budget may be reduced so much that it will be effectively impossible to reject any more null hypotheses.

Although these policies may appear wasteful as there is no reward for wealth which has not been invested, they are aimed to preserve some of their current budget for future tests in case the hypotheses considered in the beginning of the testing procedure are not particularly trustworthy.

This investing rule is therefore particular suited for scenarios were the total number of false discoveries in long exploration sessions, potentially across multiple users, should be controlled. 

\subsection{$\gamma$-Fixed Investing Rule}

A different \emph{non-thrifty} procedure assigns to each hypothesis the same budget $\alpha^*$. In particular, we call $\gamma$\emph{-fixed} a procedure that assigns to each null hypothesis a fixed budget $\alpha_j$ equal to a fraction of the initial $\alpha$-wealth $W(0)$, that is $\alpha^* = W(0)/\left(W(0)+\gamma\right)$, as long as any $\alpha$-wealth is available. 

The details of the $\gamma$-fixed procedure controlling $mFDR_{\eta}$ at level $\alpha$ can be found in the procedure for Investing Rule \ref{alg:gamma}.

\begin{algorithm}
\scriptsize
    \caption{$\gamma$-fixed}
    \label{alg:gamma}
    \begin{algorithmic}[1]
        \State $W(0) = \eta \alpha$
        \State $\alpha^* = \frac{W(0)}{\gamma + W(0)}$
        \While {$W(j-1) - \frac{\alpha^*}{1 - \alpha^*} \ge 0 \text{, for } j=1,2,\ldots$}
            \If {$p(H_j) < \alpha^*$}
                \State $W(j) = W(j - 1) + \omega$
            \Else
                \State $W(j) = W(j - 1) - \frac{\alpha^*}{1 - \alpha^*} = W(j - 1) - \frac{W(0)}{\gamma}$
            \EndIf
        \EndWhile
    \end{algorithmic}
\end{algorithm}

Note that we define $\alpha^*$ as $W(0)/(\gamma + W(0))$ to ensure that the subtraction of the wealth is constantly $W(0)/\gamma$.
Different choices for the parameter $\gamma$ characterize how conservative the investing policy is. If there is high confidence on the first observed hypotheses being actual discoveries small values of $\gamma$ (i.e. 5,10,20) would make more sense. Vice versa a high value of $\gamma$ ensures that even if the first hypotheses are true null, a large part of the $\alpha$ wealth is preserved. Good choices for that setting would be $\gamma = 50, 100$.

\subsection{$\delta$-Hopeful Investing Rule}

In a slight  variation of the $\gamma$-fixed investing rule, we say that a policy is $\delta$-hopeful if the budget is assigned to each hypothesis ``\emph{hoping}'' that at least one of the next $\delta$ hypotheses will be rejected. Each time a null hypothesis is rejected the budget obtained from the rejection is re-invested when assigning budget over the next $\delta$ null hypotheses. $\gamma$-fixed and $\delta$-hopeful operate by spreading the amount of $\alpha$-wealth over a fixed number of hypotheses (either $\gamma$ or $\delta$),  $\delta$-hopeful is however ``\emph{less conservative}'' than $\gamma$-fixed as it always operates by investing \emph{all} currently available $\alpha$-wealth over the next $\delta$ hypotheses. 
So it is a much more optimistic procedure, which works well if most alternative hypotheses are expected to accepted.  
The details of the $\delta$-fixed procedure controlling $mFDR_{\eta}$ at level $\alpha$ can be found in the procedure for Investing Rule \ref{alg:delta}.

\begin{algorithm}
\scriptsize
    \caption{$\delta$-hopeful}
    \label{alg:delta}
    \begin{algorithmic}[1]
        \State $W(0) = \eta \alpha$
        \State $\alpha^* = \frac{W(0)}{\delta + W(0)}$
        \State $k^* = 0$
        \While{$W(j - 1) - \frac{\alpha^*}{1 - \alpha^*} \ge 0 \text{, for } j = 1, 2, \ldots$}
            \If {$p(H_j) < \alpha^*$}
                \State $W(j) = W(j - 1) + \omega$
                \State $\alpha^* = \min\left(\alpha, \frac{W(j)}{\delta + W(j)}\right)$
                \State $k^* = j$
            \Else
                \State $W(j) = W(j - 1) - \frac{\alpha^*}{1 - \alpha^*} = W(j - 1) - \frac{W(k^*)}{\alpha^*}$
            \EndIf            
        \EndWhile
    \end{algorithmic}
\end{algorithm}

\subsection{$\epsilon$-Hybrid Investing Rule}

Because \ainv allows contextual information to be incorporated, the power of the resulting procedure is related to how well the design heuristic fits the actual data exploration scenario.  For example, when the data  exhibits  more randomness, the $\gamma$-fixed rule 
tends to have more power than the $\delta$-hopeful rule.  Intuitively, the $\alpha$-wealth decreases when testing a true null hypothesis, because the expectation of the change of wealth is negative when the $p$-value is uniformly distributed on $[0,1]$. Thus the initial $\alpha$-wealth is on average larger than the $\alpha$-wealth available at subsequent steps. Furthermore, since the $\gamma$-fixed rule invests a constant fraction of the initial wealth, the power tends to be larger than $\delta$-hopeful.  

On the contrary, when the data is less random, the $\gamma$-fixed rule becomes less powerful than $\delta$-hopeful rule. 
The reason is that in this setting more significant discoveries tend to keep the subsequent $\alpha$-wealth high, potentially even higher than the initial wealth. We study this difference in more detail in Section \ref{sec:eval}.

In order to have a robust performance in terms of power and false discovery rate, we design $\epsilon$-hybrid investing rule that adjust the $\alpha_j$ assigned to the various tests based on the estimated data randomness.  
Our estimation of the randomness of the data is based on the ratio of rejected null hypothesis  over a sliding window $H_d$ constituted by the last $d$ null hypotheses observed on a stream. We then compare this ration with a ``\emph{randomness threshold}'' $\epsilon\in(0,1)$  and we conclude whether the data exhibits high randomness or not. The procedures is outlined in Investing Rule \ref{alg:epsilon}.

\begin{algorithm}
\scriptsize
    \caption{$\epsilon$-hybrid}
    \label{alg:epsilon}
    \begin{algorithmic}[1]
        \State $W(0) = \eta \alpha$
        \State $k^* = 0$
        \State $H_d = []$ // Sliding window of size $d$
        \While {$W(j-1) > 0 \text{, for } j=1,2,\ldots$}
            \If {Rejected($H_d$) $\le$ $|H_d|\epsilon$}
                \State $\alpha_j = \frac{W(0)}{\gamma + W(0)}$
            \Else
                \State $\alpha_j = \min\left(\alpha, \frac{W(k^*)}{\delta + W(k^*)}\right)$
            \EndIf
            \If {$W(j-1) - \frac{\alpha_j}{1 - \alpha_j} \ge 0$}
                \If {$p(H_j) < \alpha_j$}
                    \State $W(j) = W(j - 1) + \omega$
                    \State $k* = j$
                    \State $H_d[j] = R_j = 1$
                \Else
                    \State $W(j) = W(j - 1) - \frac{\alpha_j}{1 - \alpha_j}$
                    \State $H_d[j] = R_j = 0$
                \EndIf
            \EndIf
        \EndWhile
    \end{algorithmic}
\end{algorithm}

\subsection{Investment based on Support Population}

In this section we discuss how to adjust the budget of each hypothesis according to the amount of data which is available in order to compute the $p$-value of that same hypothesis. The main intuition for this procedure is that, as it is most likely to observe high \pvals for hypotheses which rely on a small number of data points, we should should not invest as much $\alpha$-wealth on those hypotheses. In this section we discuss how to \emph{bias} the amount budget assigned to each hypothesis so that hypotheses with more support data receive more ``\emph{trust}'' (in terms of budget) from the procedure.

Let us denote as $|n|$ the total amount of data being used and by $|j|$ the available data for testing the $j$-th null hypothesis $H_t$.
A simple way of correcting the assignment of the budget $\alpha_j$ in any of the previously mentioned hypothesis is to assign to the test of the hypothesis $\alpha_j f(\frac{|j|}{|n|})$. 
Depending on the choice of $f(\cdot)$ the impact of the correction may be more or less severe. Some possible choices for $f(\cdot)$ would be $f(\frac{|t|}{|n|})= \left(\frac{|t|}{|n|}\right)^{\psi}$ for possible values of $\psi = 1, 2/3, 1/2, 1/3,\ldots$. We present an example policy based on the $\gamma$-fixed rule, the $\psi$-support rule in Investing Rule \ref{alg:psi}.

\begin{algorithm}
\scriptsize
    \caption{$\psi$-support}
    \label{alg:psi}
    \begin{algorithmic}[1]
        \State $W(0) = \eta \alpha$
        \State $\alpha^* = \frac{W(0)}{\gamma + W(0)}$
        \While {$W(j-1) > 0 \text{, for } j=1,2,\ldots$}
            \State $\alpha_j = \alpha^*\left(\frac{|t|}{|n|}\right)^{\frac{1}{2}}$
            \If {$W(j-1) - \frac{\alpha_j}{1 - \alpha_j} \ge 0$}
                \If {$p(H_j) < \alpha_j$}
                    \State $W(j) = W(j - 1) + \omega$
                \Else
                    \State $W(j) = W(j - 1) - \frac{\alpha_j}{1 - \alpha_j}$
                \EndIf
            \EndIf
        \EndWhile
    \end{algorithmic}
\end{algorithm}

\subsection{What Happens If the Wealth is 0}
\label{sec:fdrcontrol:nowealth}

Among all our proposed investing policies, only $\beta$-farsighted is ``\emph{thrifty}'',that it is never fully commits its available $\alpha$-wealth. 
Still, the available wealth for $\beta$-farsighted could eventually become extremely small, to the point that no more hypotheses can be rejected.
All the remaining procedures are ``\emph{non-thrifty}'' and can thus reach zero $\alpha$-wealth, in which case the user (theoretically) should stop exploring. 

It is only natural to wonder if it would be possible for the user to somehow ``\emph{recover}'' some of the lost $\alpha$-wealth and thus continuing the testing procedure.
One possible way to do so, would require the user to reconsider and possibly overturn some of the previous decisions on whether to reject or accept some null hypotheses using alternative testing procedures (i.e., the Benjamini-Hochberg procedure). 

There are however several challenges to be faced when pursuing this strategy: 1) great care has to be put on haw to combine results from different testing procedures (i.e., control of FDR for a subsets of hypotheses and control of \emph{mFDR} for a distinct subset of hypotheses) and 2) testing hypotheses for a second time given the outcomes of other test implies a clear (and strong) dependence between the outcome of the tests and the \pval associated with the null hypotheses being considered. 
Therefore, depending on the context such control could only be achieved  given additional assumptions about the level of control or would require adding additional data or the use of a hold-out dataset.  
We aim to study this problem in detail as part of future work. 


\begin{figure*}
\centering
\captionsetup[subfloat]{farskip=2pt,captionskip=1pt}
    \subfloat[][\scriptsize 75\% Null: Avg. Disc]{
        \includegraphics[height=2.4cm, trim=5mm 0mm 4mm 0mm]{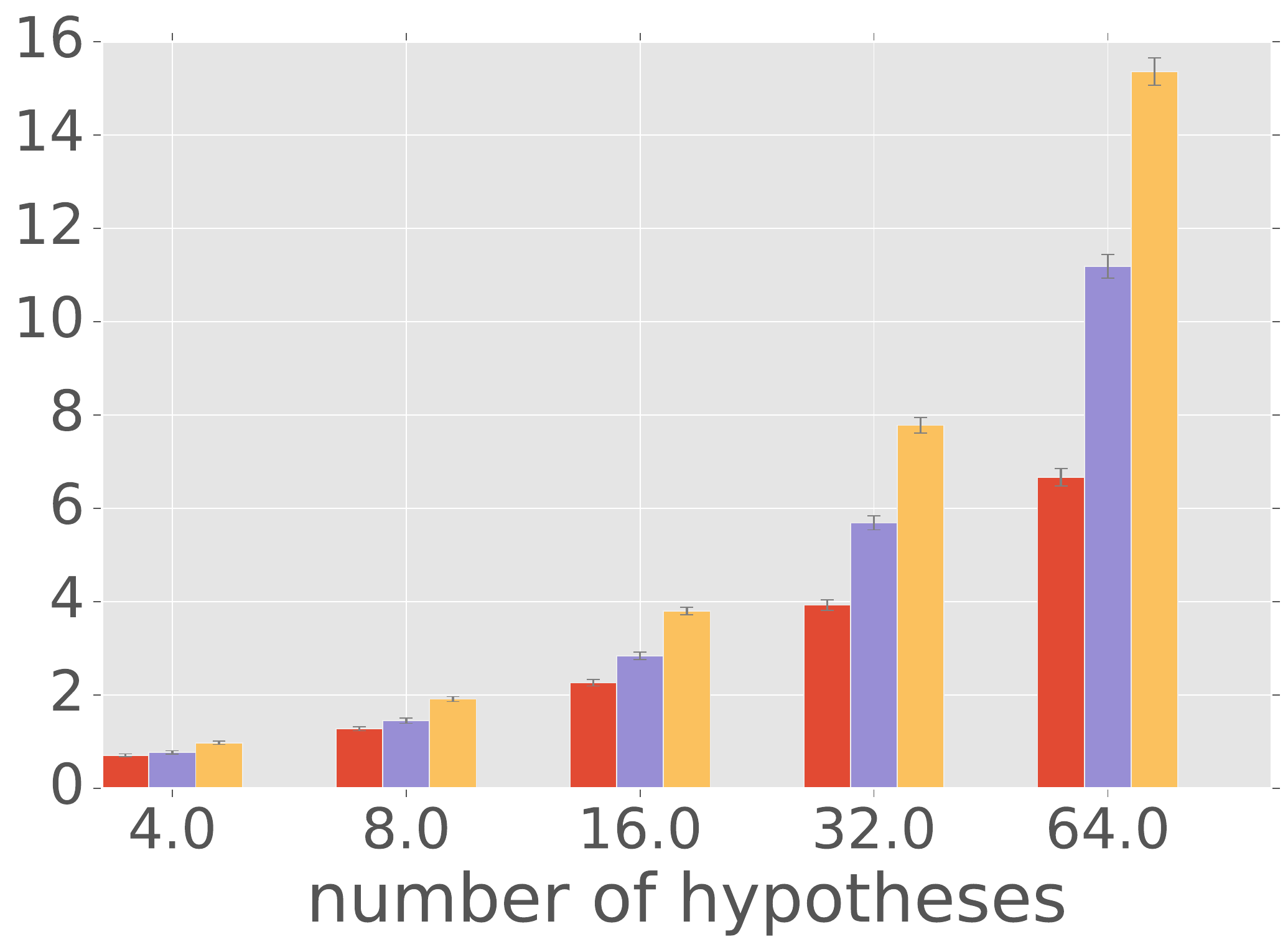}
        \label{fig:static-75-null-1}
    }\hspace*{\fill}
	\subfloat[][\scriptsize 75\% Null: Avg. FDR]{
        \centering
        \includegraphics[height=2.3cm, trim=5mm 0mm 4mm 0mm]{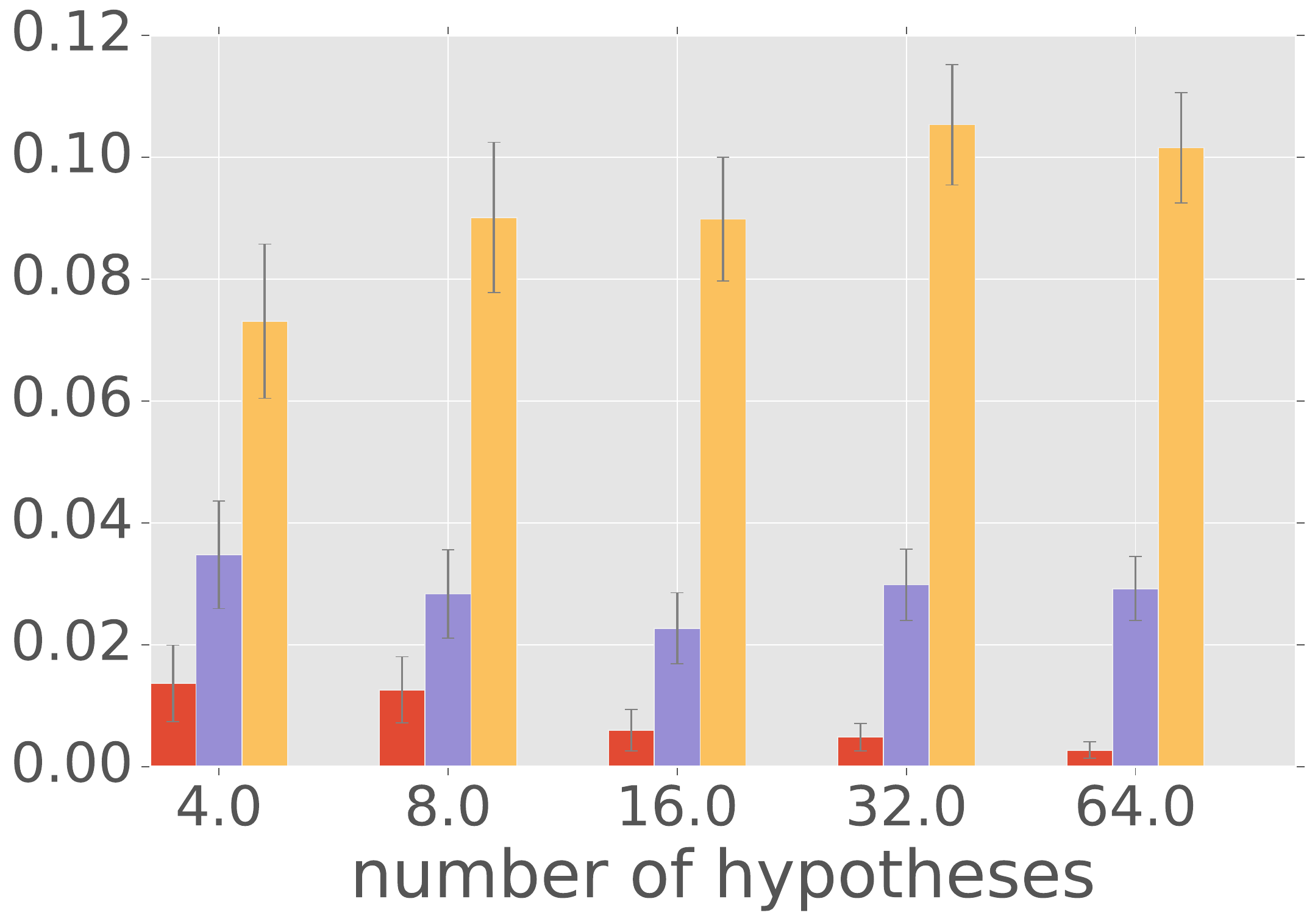}
        \label{fig:static-75-null-2}
    	}\hspace*{\fill}
		\subfloat[][\scriptsize 75\% Null: Avg. Power]{
        \centering
        \includegraphics[height=2.3cm, trim=5mm 0mm 4mm 0mm]{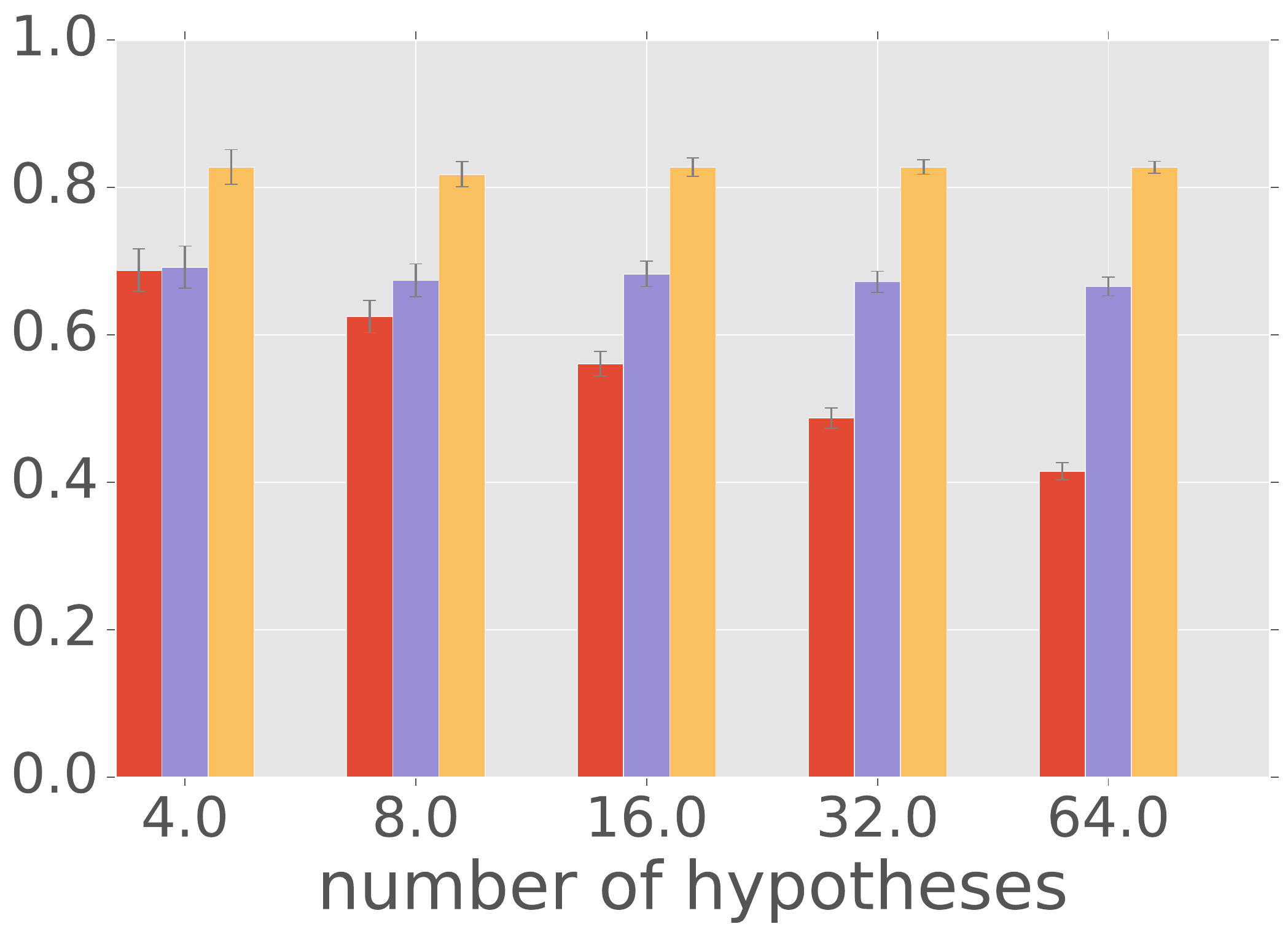}
        \label{fig:static-75-null-3}
    	}\hspace*{\fill}
		\subfloat[][\scriptsize 100\% Null: Avg. Disc]{
        \centering
        \includegraphics[height=2.3cm, trim=5mm 0mm 4mm 0mm]{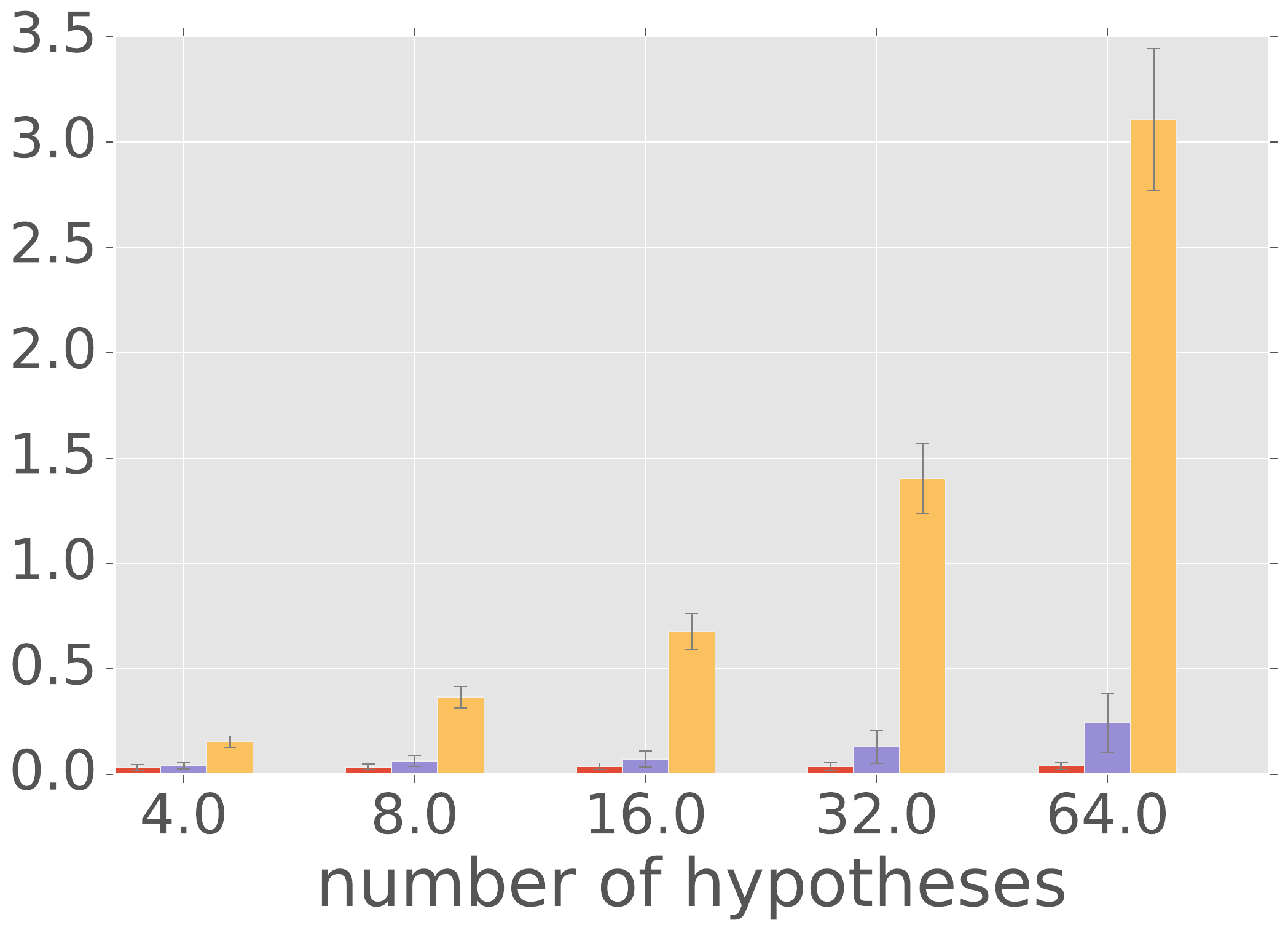}
        \label{fig:static-100-null-1}
    	}\hspace*{\fill}
		\subfloat[][\scriptsize 100\% Null: Avg. FDR]{
        \centering
        \includegraphics[height=2.3cm, trim=5mm 0mm 4mm 0mm]{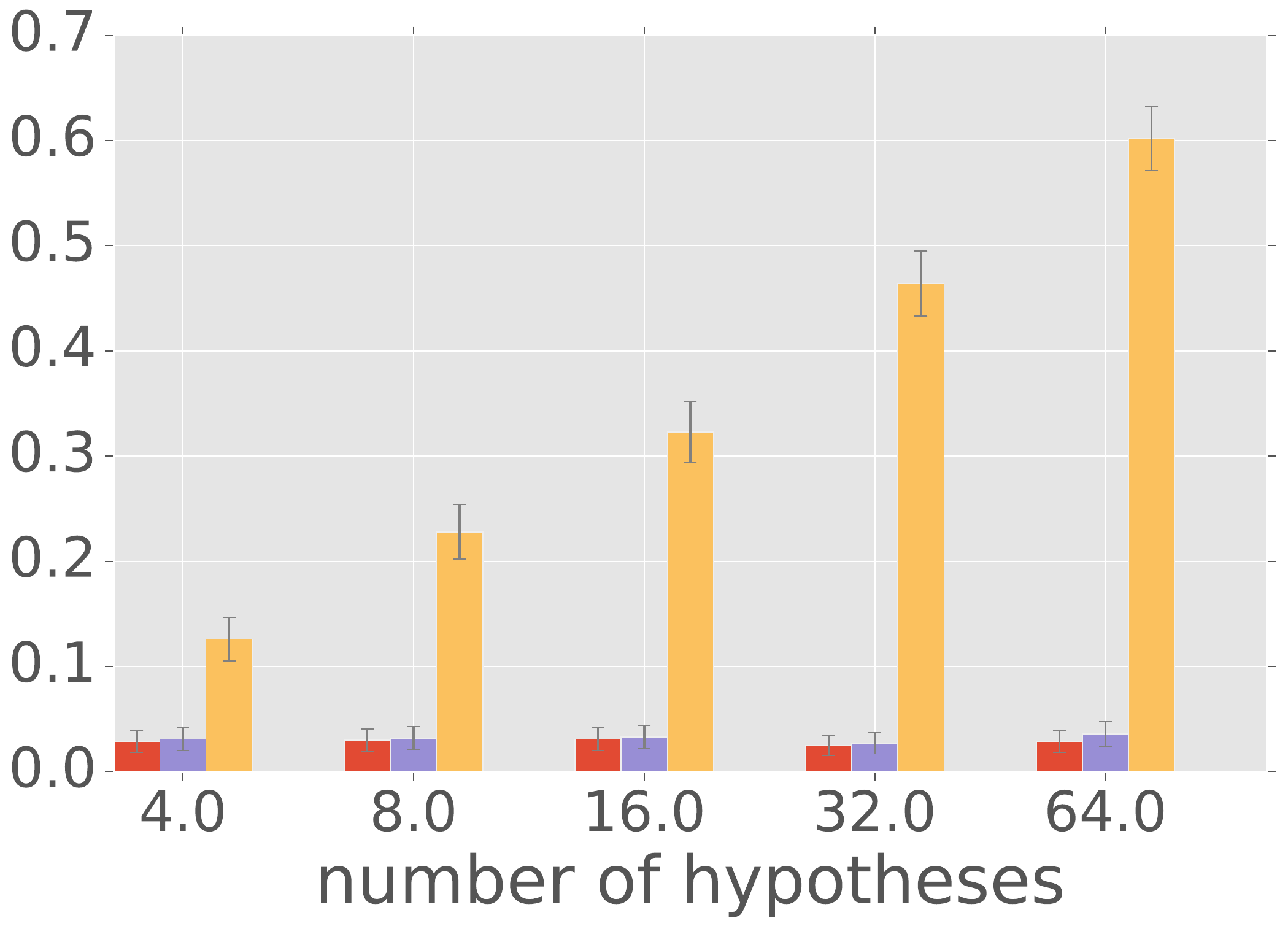}
        \label{fig:static-100-null-2}
    	}\\
		\subfloat{
        \centering
        \includegraphics[width=0.3\textwidth]{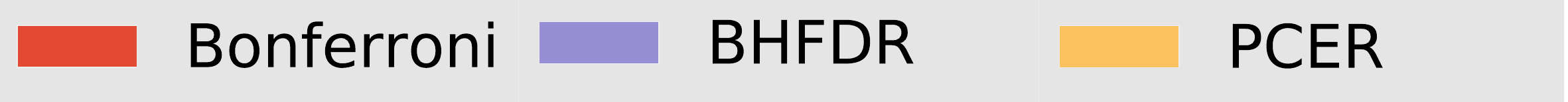}
    }
  
    \caption{Exp.1a: Static Procedures on Synthetic Data}
    \label{fig:exp-1a}
\end{figure*}



\section{The Most Important Discoveries}
\label{sec:fdrcontrol:important}

In Section~\ref{sec:ui} we argued, that the user should be able to \emph{mark} the important hypotheses (e.g., the ones she wants to include in a publication). 
This is particularly important as \system{} uses default hypotheses, which the user might consider as less important. 
In the following we show that if these ``\emph{important discoveries}'' are selected from all the discoveries given by a testing procedure that controls FDR at level $\alpha$ independently of their \pvals, then the FDR for the set of important discoveries is controlled at level $\alpha$ as well. 

\vspace{-1.5ex}
\begin{theorem}\label{thm:impdisc}
Assume that we executed a collection of hypothesis tests with a rejection rule that controls the FDR at $\alpha$. Assume that the procedure rejected the set of null hypotheses
$R=\{R_1,\dots,R_r\}$, and let $V\subseteq V$ be the set of false discoveries. If the null hypothesis tests are independent then
 for any subset $R'\subseteq R$ we have $E[|V\cap R'|/||R']\leq \alpha$.
\end{theorem}
\vspace{-1.5ex}

\begin{proof}
Let $p_1,\dots,p_{|R|}$ be the $p$-values of the rejected hypotheses. Since the rejection rule controls the FDR at $\alpha$ we have

\vspace{-2.5ex}\begin{equation}\scriptsize
\label{eq:sfdr1}
 \sum\limits_{i=1}^{|R|}\frac{i}{|R|} P(|V|=i~|~P_1=p_1,\dots, P_r =p_r) = \alpha 
\end{equation}\vspace{-2.5ex}

Assume that $|V|=i$. A priori, the $p$-values of null hypotheses are i.i.d. uniformly distributed in $[0,1]$~\cite{}. Subject to $P_1=p_1,\dots, P_r =p_r$, the set of the $i$ null hypotheses' $p$-values is uniformly distributed among all the $i$ subsets of  the $r$ value $\{p_1,\dots, p_r\}$. Let
$p'_1,\dots,p'_{|R'|}$ be the $p$-values of the set of hypotheses $R'$, and let $p^V_i, \dots p^V_{|V|}$ be the $p$-values of the rejected null hypotheses, then 

\vspace{-2.5ex}\begin{equation}\scriptsize
\label{eq:sfdr2}
\begin{split}
E[|V\cap R'|~|~|V|=i]= \\ E[|\{p'_1,\dots,p'_{|R'|}\}\cap \{p^V_1\dots P^V_{|V|} \}|~|~|V|=i]= i \frac{|R'|}{|R|}.
\end{split}
\end{equation}

Combining equations (\ref{eq:sfdr1}) and (\ref{eq:sfdr2}) we get:

\vspace{-6.5ex}\begin{equation}
\scriptsize
\begin{split}
E \left [\frac{|V\cap R'|}{|R'|} \right ] = \\
 \sum_{i=1}^{|R|}E\left [ \frac{|V\cap R'|}{|R'|}~|~|V|=i\right]P(|V|=i~|~P_1=p_1, \dots, P_r =p_r) \\
 = \sum_{i=1}^{|R|}\frac{1}{|R'|}i \frac{|R'|}{|R|}P(|V|=i~|~P_1=p_1,\dots, P_r =p_r) = \alpha
\end{split}
\end{equation}\vspace{-2.5ex}
\end{proof}

Consider a set  $R'$ of important discoveries selected independently of the \pvals of the corresponding null-hypothesis from a larger set of discoveries $R$ for which then $mFDR$ is controlled at level $\alpha$. Using a proof similar to the one discussed in Theorem~\ref{thm:impdisc}  it is possible to show that the $mFDR$ of $R'$ is controlled at level $\alpha$ as well.
This is an important result, as it implies that the user can select the  important discoveries from a larger pool of discoveries while maintaining the control of FDR (or mFDR) at level $\alpha$. 





\section{Experimental Evaluation}
\label{sec:eval}
In this section, we evaluate the \ainv rules in different data exploration settings to answer the following questions: 
\vspace{-1.5ex} 
\begin{enumerate*}
\item How do our \ainv rules compare to Sequential FDR?
\item What is the average power (the proportion of truly significant discoveries that are correctly identified)?
\item What is the average false discovery rate?
\end{enumerate*}
\vspace{-1.5ex} 

\begin{figure*}
\centering
\captionsetup[subfloat]{farskip=2pt,captionskip=1pt}
    \subfloat[][\scriptsize 25\% Null: Avg. Discoveries]{
        \includegraphics[width=0.22\textwidth]{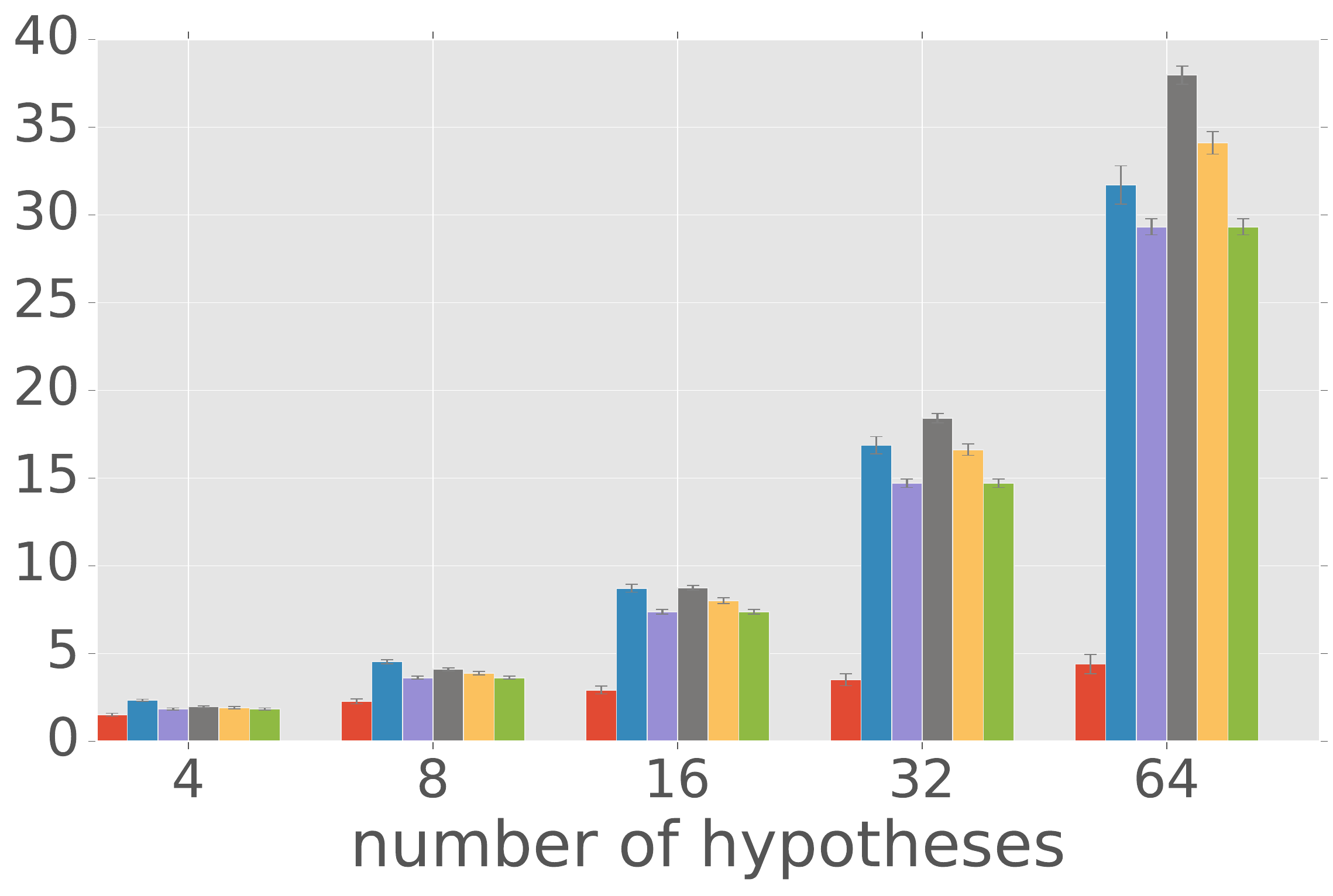}
        \label{fig:incremental-25-null-1}
    }\hspace*{\fill}
	\subfloat[][\scriptsize 25\% Null: Avg. FDR]{
        \includegraphics[width=0.22\textwidth]{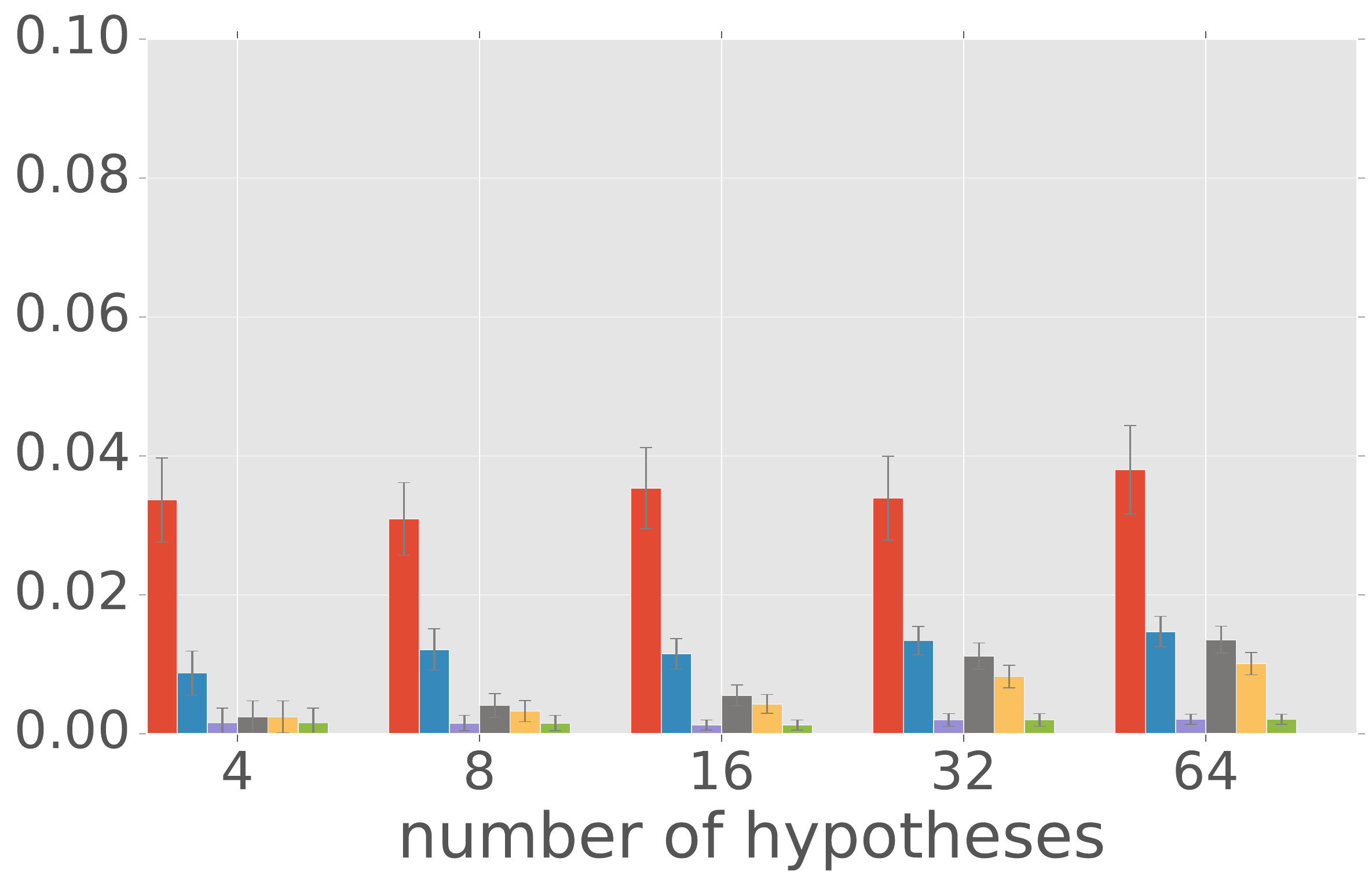}
        \label{fig:incremental-25-null-2}
    }\hspace*{\fill}
	\subfloat[][\scriptsize 25\% Null: Avg. Power]{
        \includegraphics[width=0.22\textwidth]{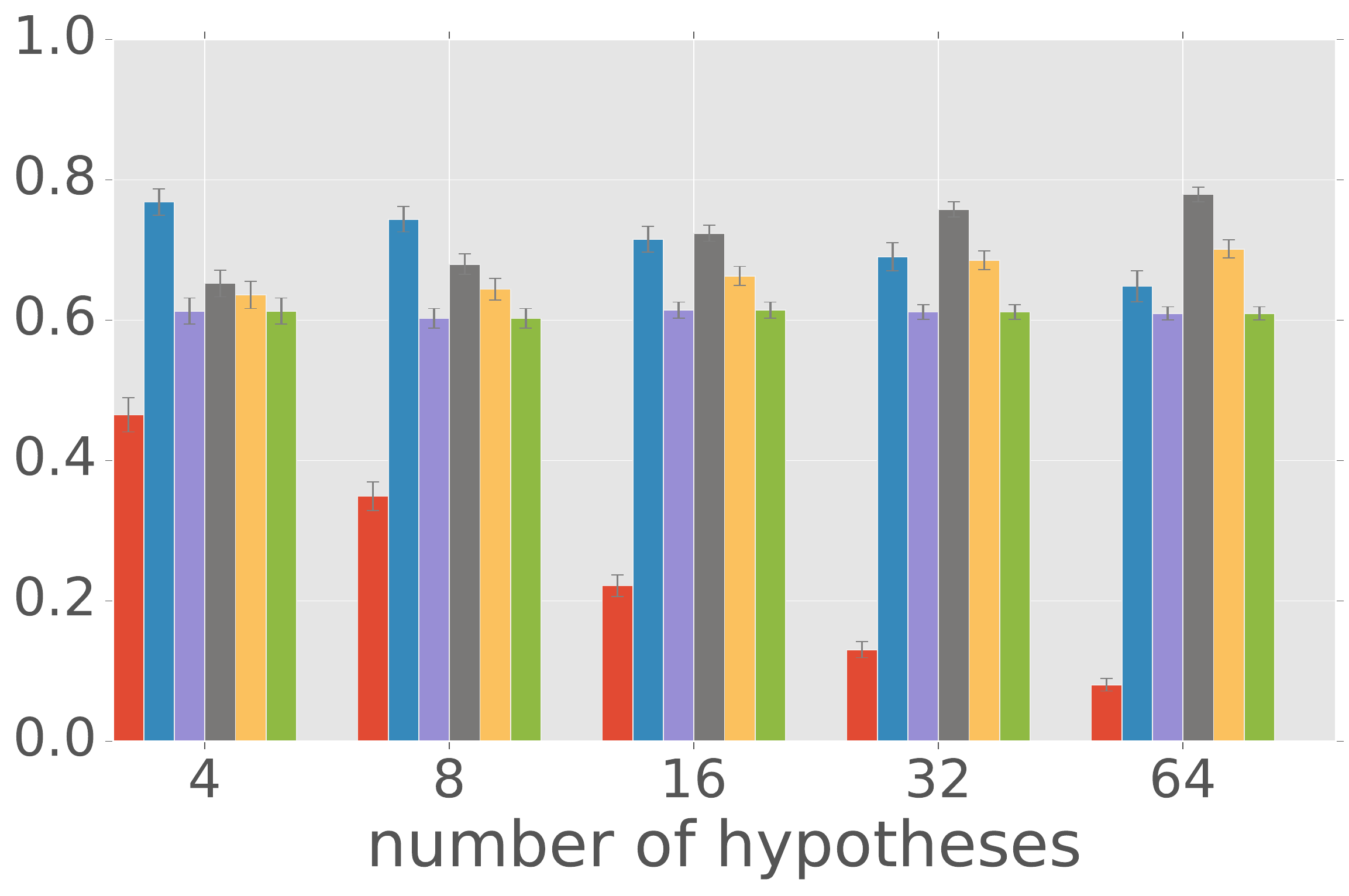}
        \label{fig:incremental-25-null-3}
    }\hspace*{\fill}
	\subfloat{
        \includegraphics[width=1.35pt]{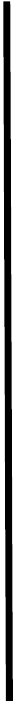}
    }\hspace*{\fill}
	\subfloat[][\scriptsize 100\% Null: Avg. Discoveries]{
	 	\setcounter{subfigure}{7}
     \includegraphics[width=0.22\textwidth]{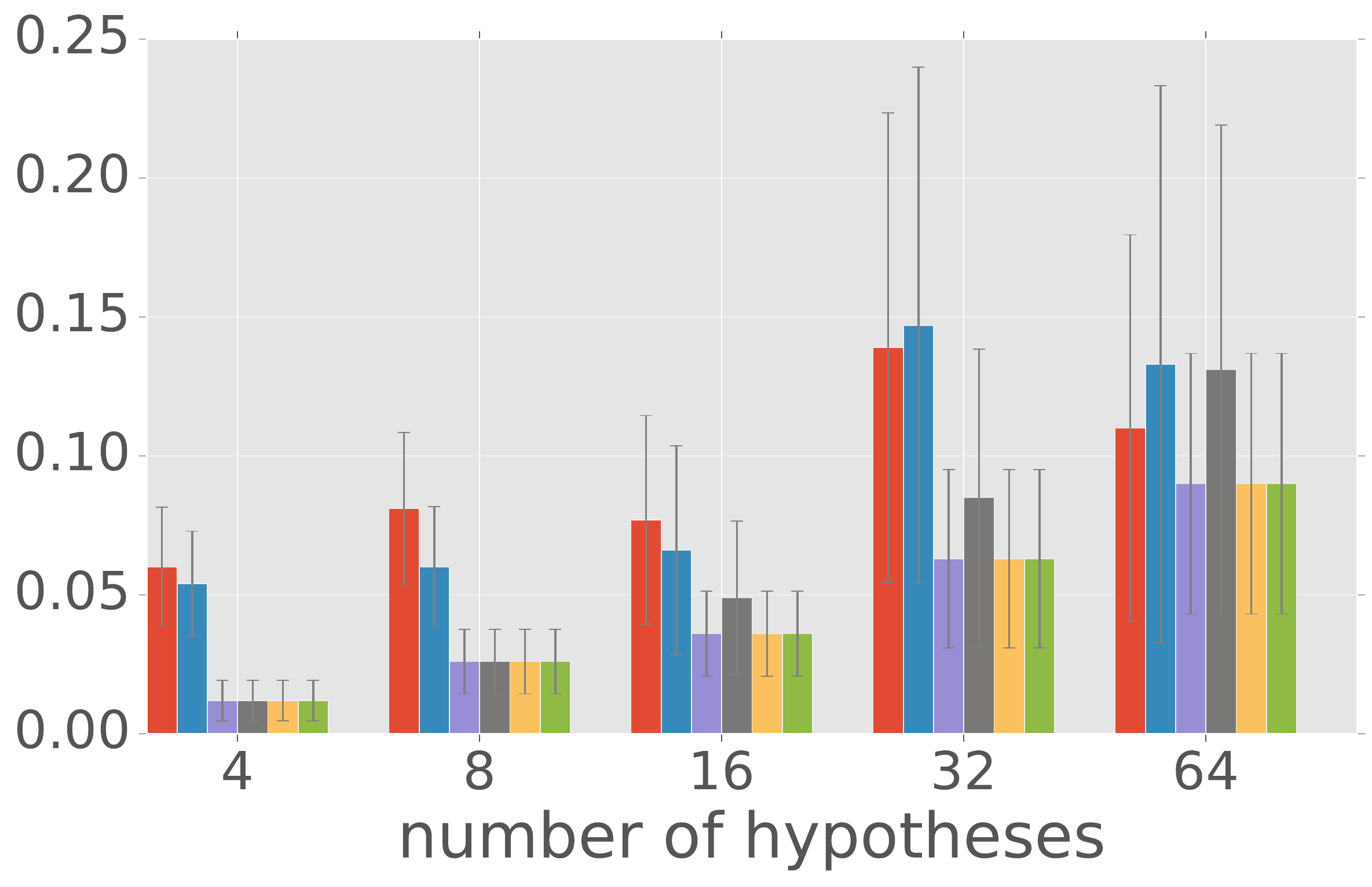}
 		\label{fig:incremental-100-null-1}
    }\hspace*{\fill}
\\
   	\subfloat[][\scriptsize 75\% Null: Avg. Discoveries]{
		\setcounter{subfigure}{4}
        \includegraphics[width=0.22\textwidth]{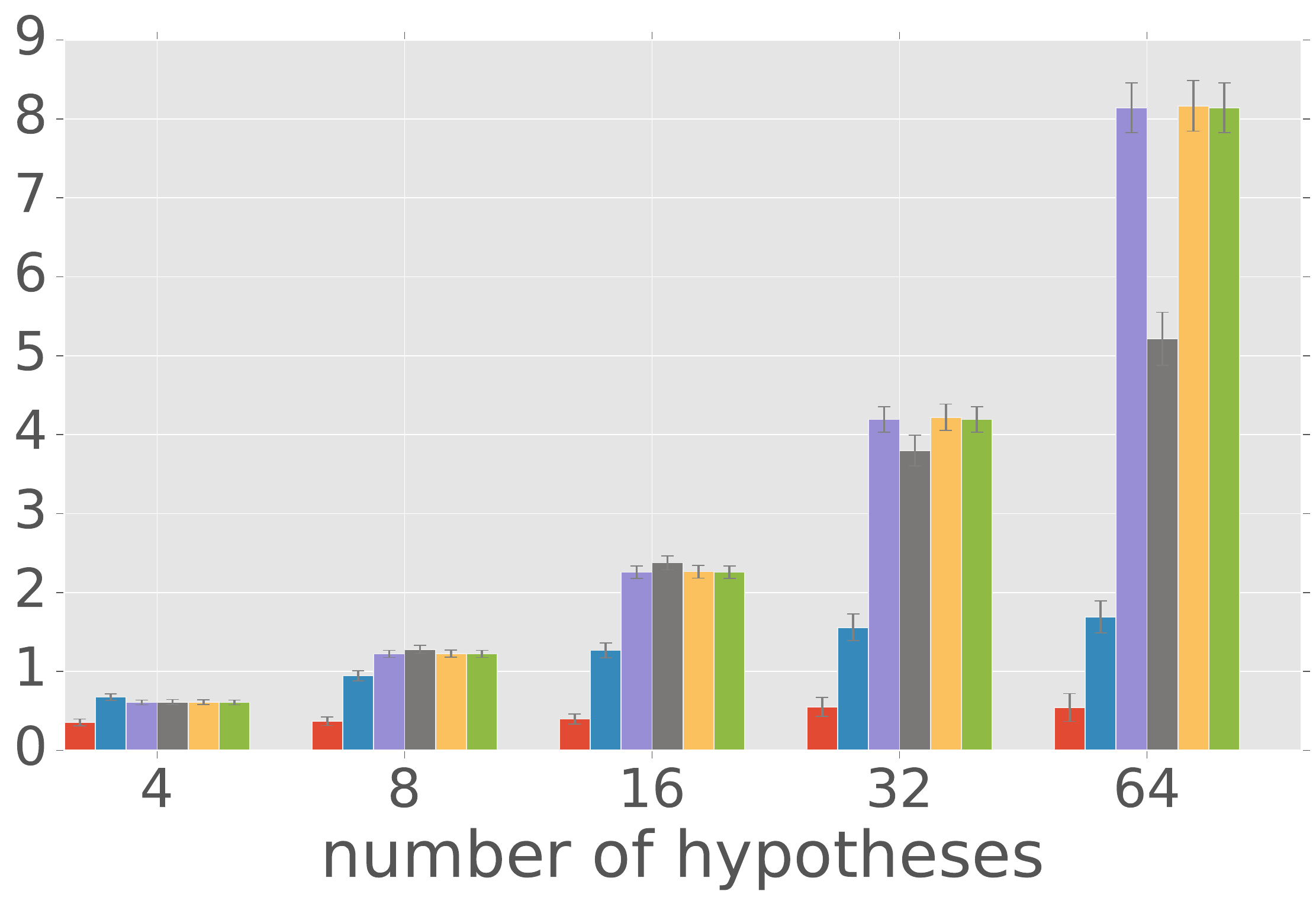}
       \label{fig:incremental-75-null-1}
    }\hspace*{\fill}
	\subfloat[][\scriptsize 75\% Null: Avg. FDR]{
       	\includegraphics[width=0.22\textwidth]{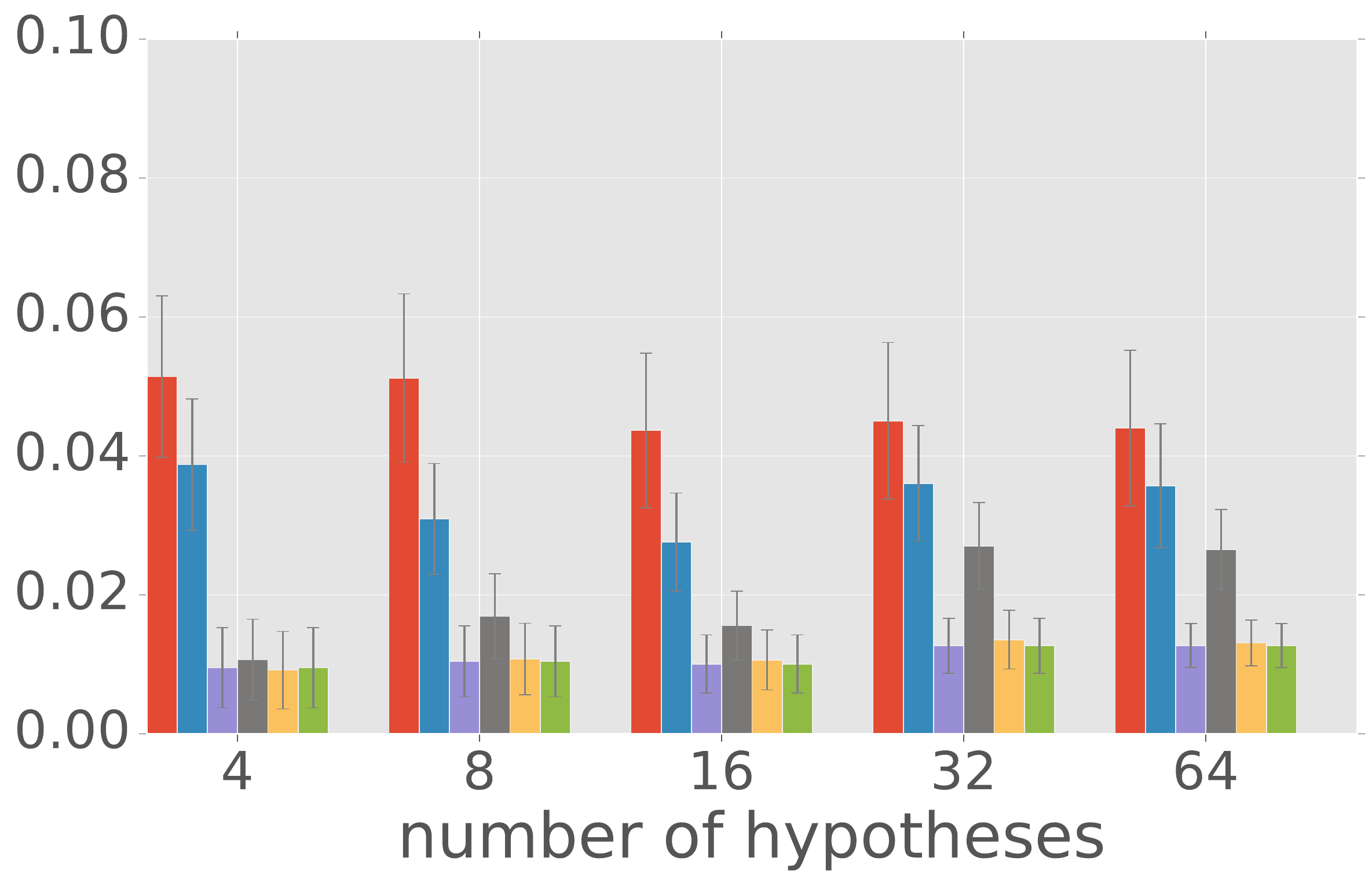}
       \label{fig:incremental-75-null-2}
    }\hspace*{\fill}
	\subfloat[][\scriptsize 75\% Null: Avg. Power]{
       	\includegraphics[width=0.22\textwidth]{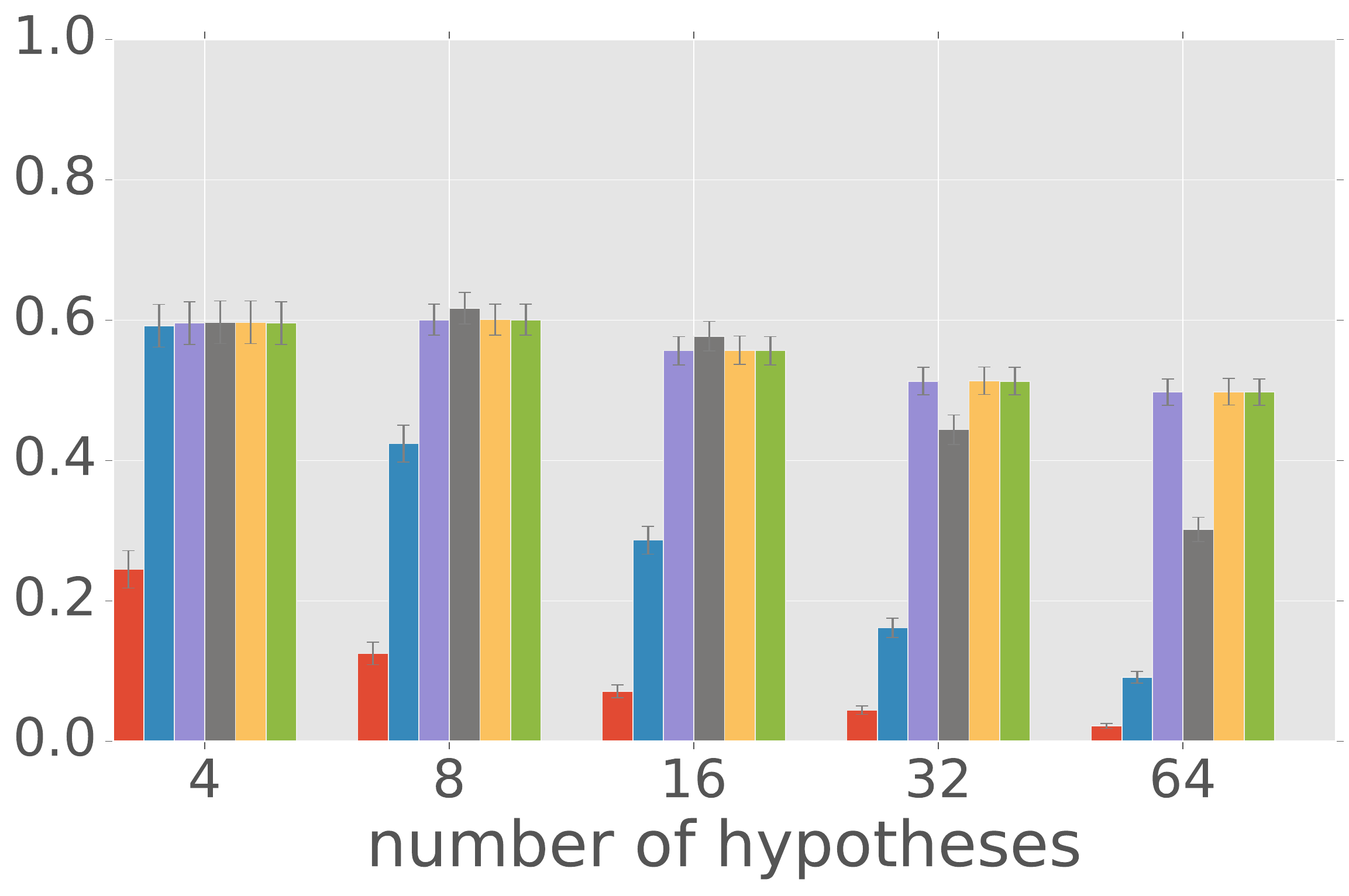}
        \label{fig:incremental-75-null-3}
    }\hspace*{\fill}
	\subfloat{
       	\includegraphics[width=1.35pt]{figures/line.pdf}
    }\hspace*{\fill}
	\subfloat[][\scriptsize 100\% Null: Avg. FDR]{
       	\includegraphics[width=0.22\textwidth]{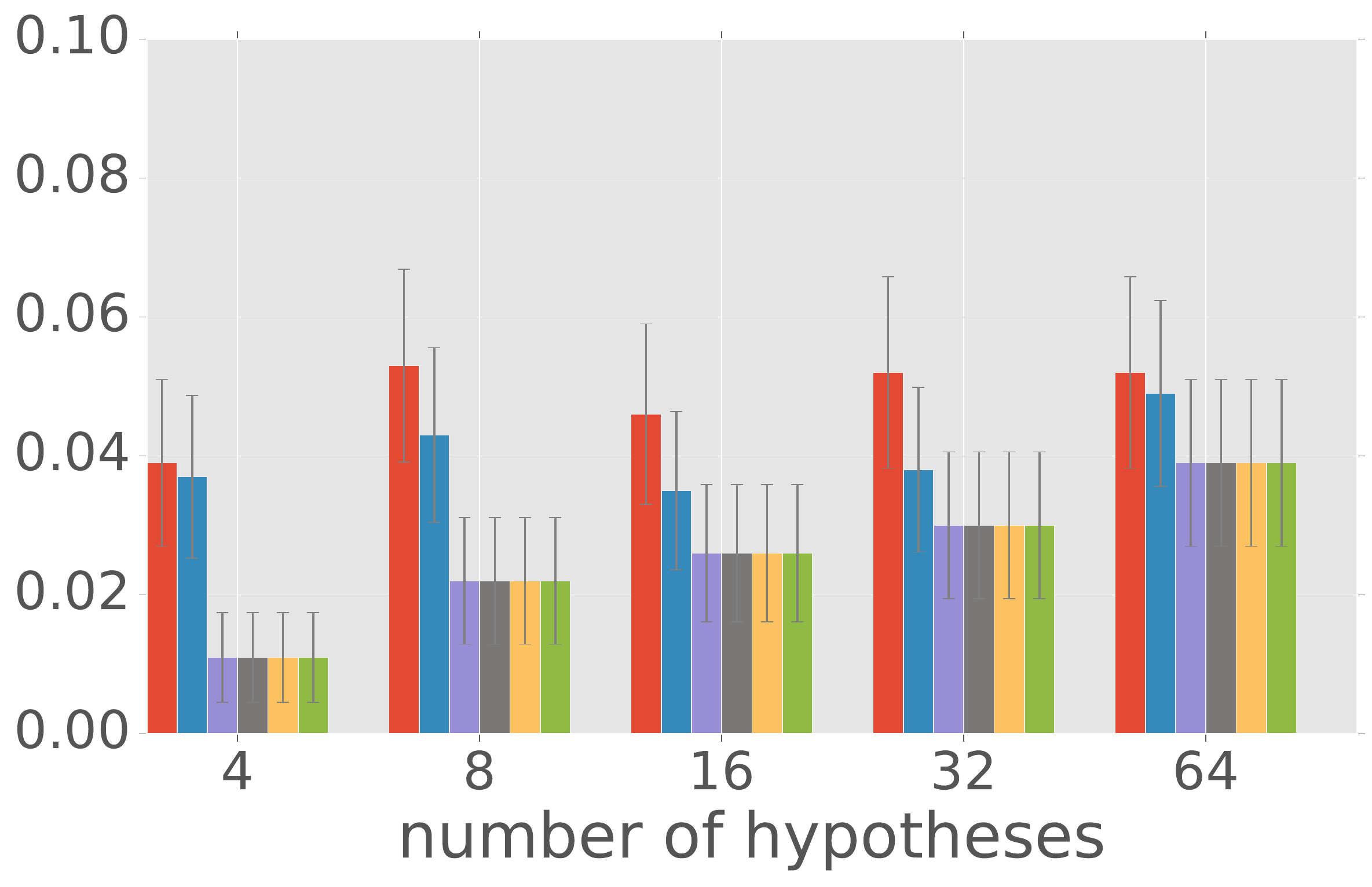}
        \label{fig:incremental-100-null-2}
    }
\\
    
    \subfloat{
        \includegraphics[width=0.7\textwidth]{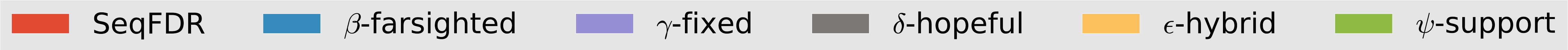}
    }
    
    \caption{Exp.1b: Incremental Procedures on Synthetic Data / Varying Number of Hypotheses}
    \label{fig:exp-1b}
\end{figure*}

\textbf{Workload/Data:} We first conduct the simulation analysis on synthetic data, and then run user-study workflows on a real-world dataset.
The statistics community considers the simulation analysis on synthetic data to be the statistically sound methodology to evaluate a multiple hypothesis testing procedure (see for example \cite{BenjaminiH95,benjamini1995controlling}), because on real-world datasets and workflows the proportion and signal-to-noise ratio of truly significant and insignificant hypotheses are hard to determine and control.

\textbf{Implementations and Setup:}
The procedures for all experiments are: 
(1) No multiple hypothesis control: Per-Comparison Error Rate (PCER) \cite{benjamini1995controlling},
(2) Static: Bonferroni Correction (Bonferroni) \cite{bonferroni1936teoria} and Benjamini-Hochberg (BHFDR) \cite{benjamini1995controlling}
(3) Incremental but non-interactive: Sequential FDR (SeqFDR) \cite{GSell}  
(4) Incremental and interactive: \ainv rules of this paper. 

We modified our system to also execute static procedures. We emphasize that the static-versus-incremental comparison only serves as a reference as the static procedures are essentially not suitable for data exploration as discussed in Section \ref{sec:methods}.

For all configurations, we set $\alpha$ to $0.05$ and estimate the average false discoveries, the average FDR (i.e., the average of the ratios of the false discoveries over all discoveries), and the average power and their corresponding 95\% confidence intervals.

\begin{figure*}
\centering
\captionsetup[subfloat]{farskip=2pt,captionskip=1pt}
    \subfloat[][\scriptsize 25\% Null: Avg. Discoveries]{
       \includegraphics[height=1.8cm]{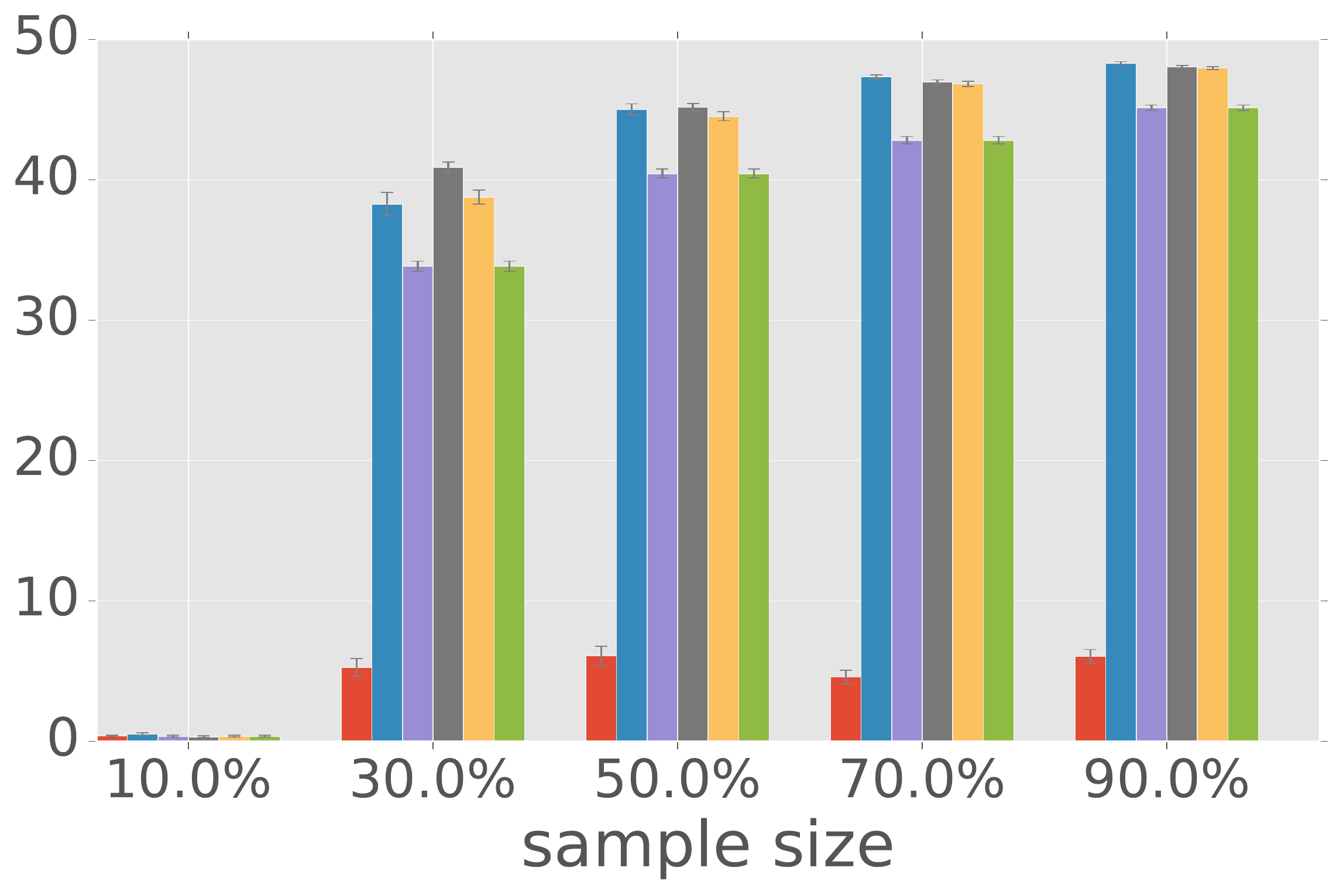}
	        \label{fig:incremental-n20-25-null-1}
    }\hspace*{\fill}
	\subfloat[][\scriptsize 25\% Null: Avg. FDR]{
       	\includegraphics[height=1.8cm]{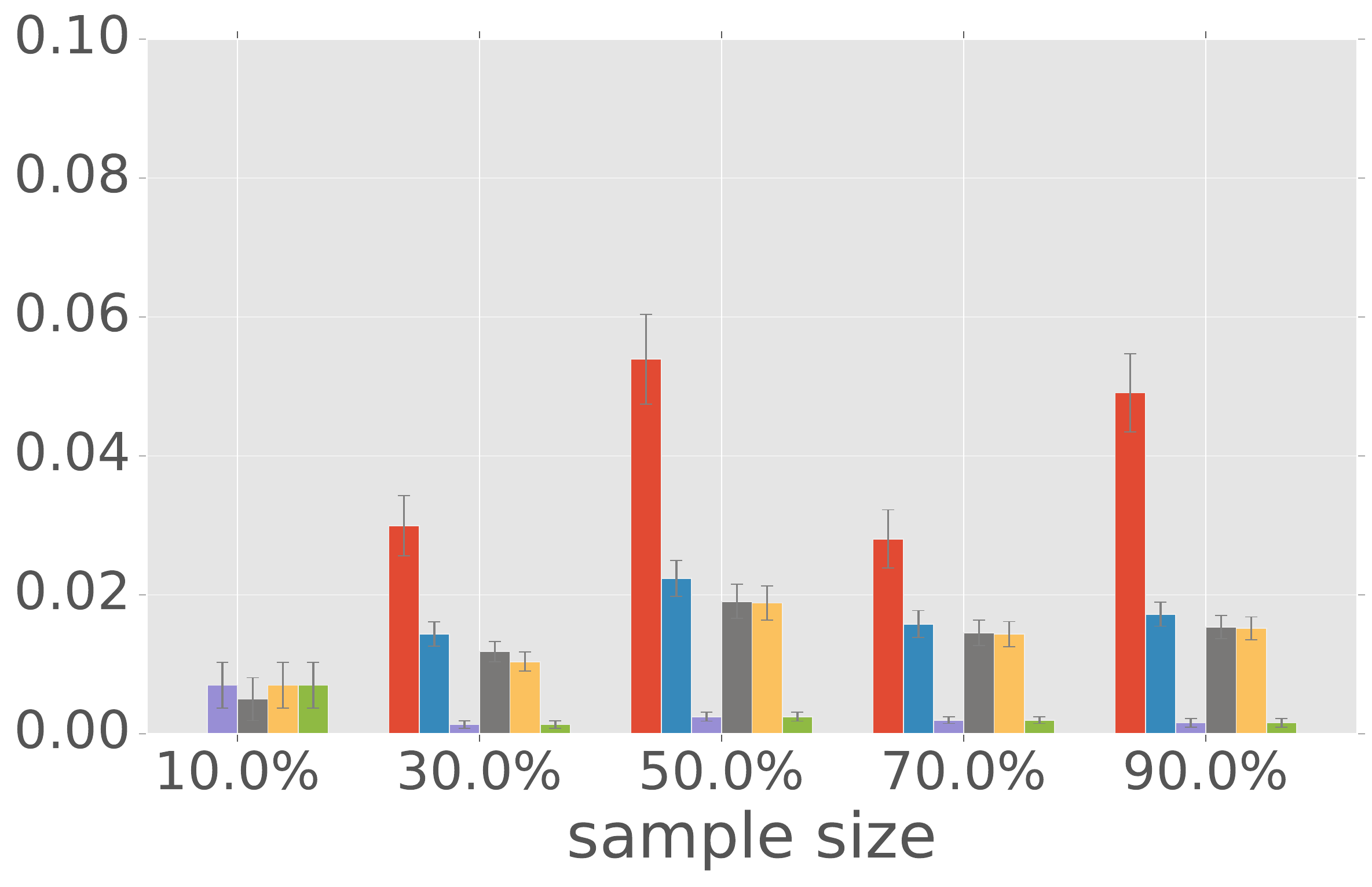}
       \label{fig:incremental-n20-25-null-2}
    }\hspace*{\fill}
 	\subfloat[][\scriptsize 25\% Null: Avg. Power]{
       \includegraphics[height=1.8cm]{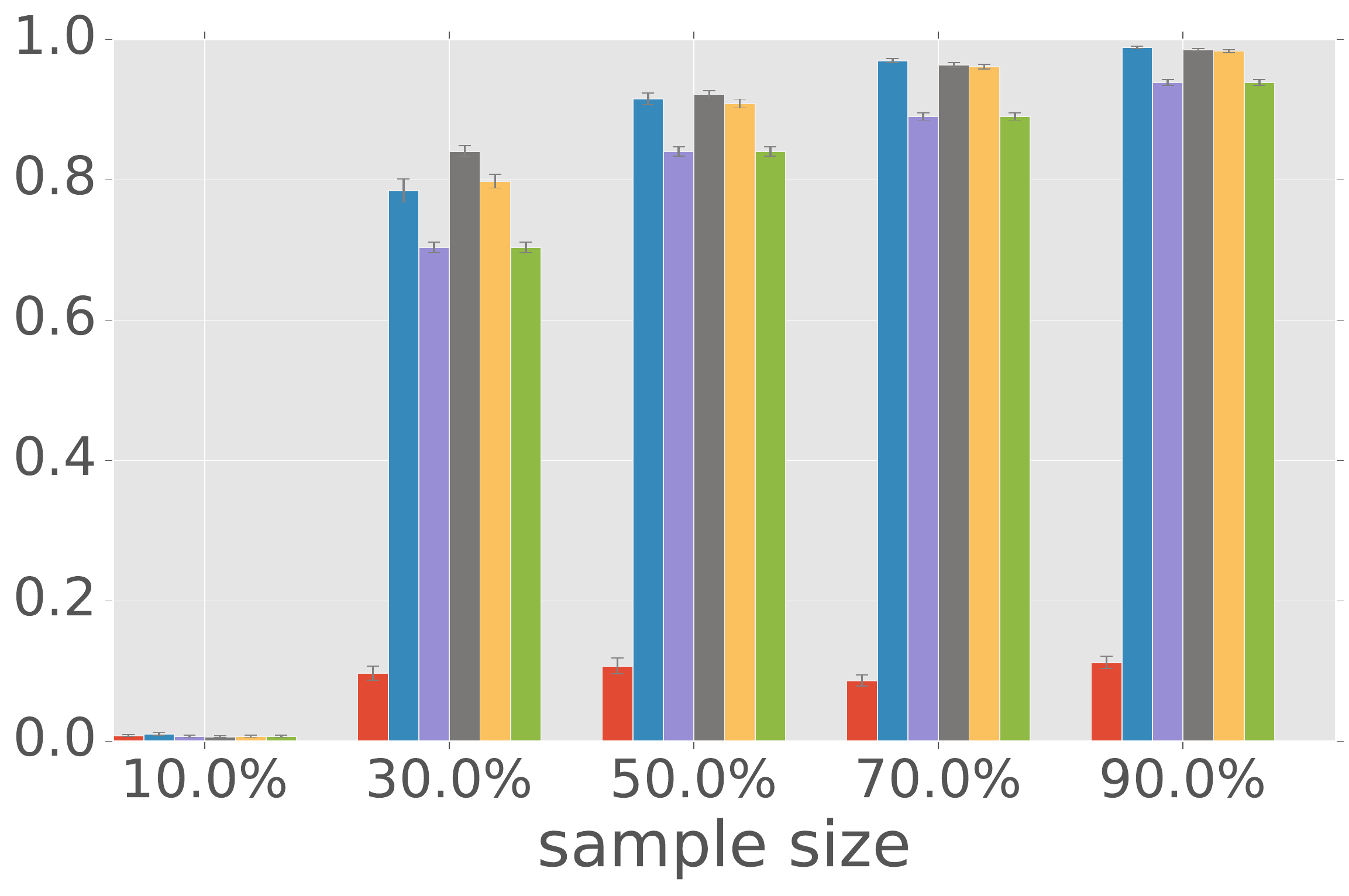}
       \label{fig:incremental-n20-25-null-3}
    }\hspace*{\fill}
	\subfloat[][\scriptsize 75\% Null: Avg. Discoveries]{
       \includegraphics[height=1.8cm]{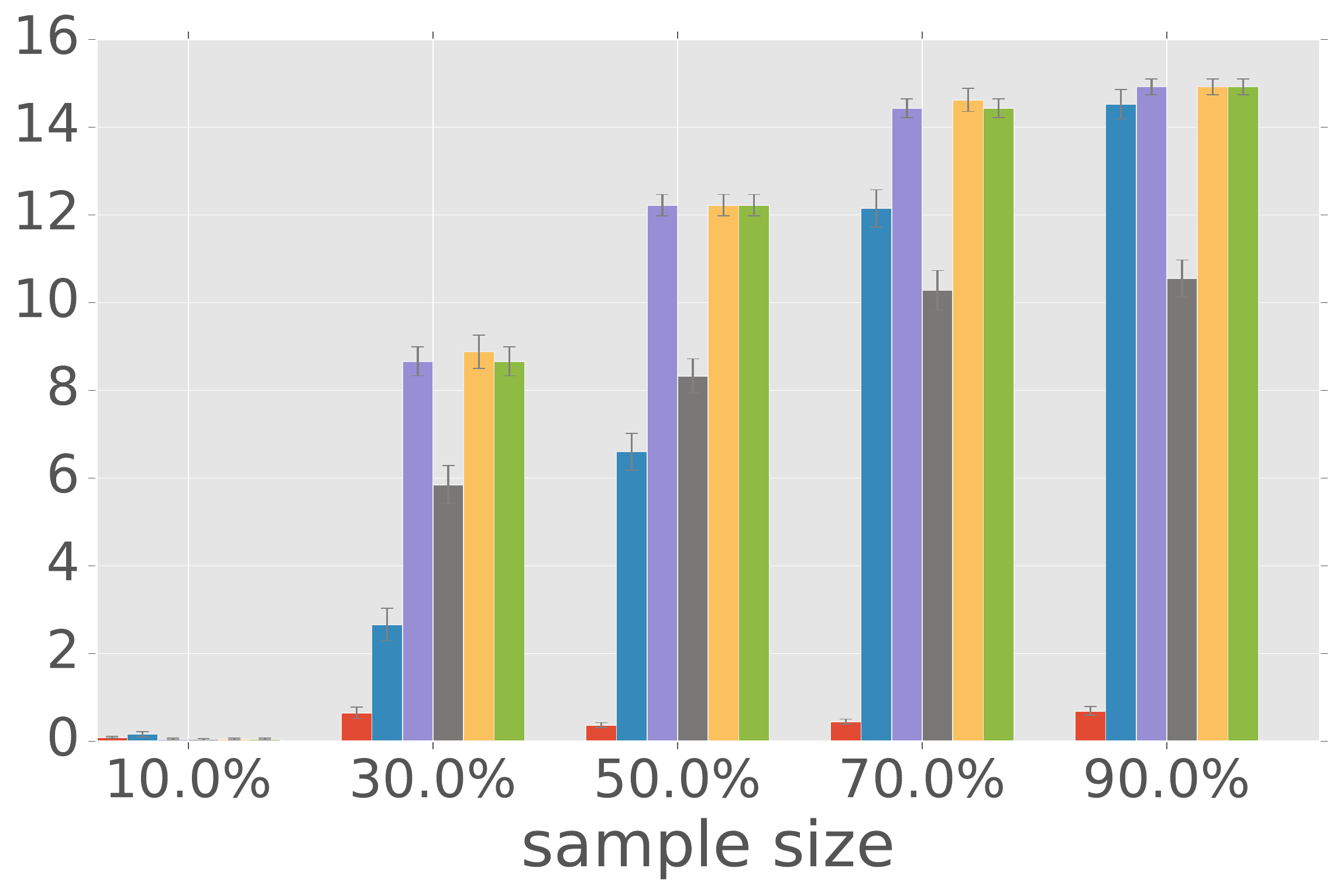}
       \label{fig:incremental-n20-75-null-1}
    }\hspace*{\fill}
	\subfloat[][\scriptsize 75\% Null: Avg. FDR]{
       \includegraphics[height=1.8cm]{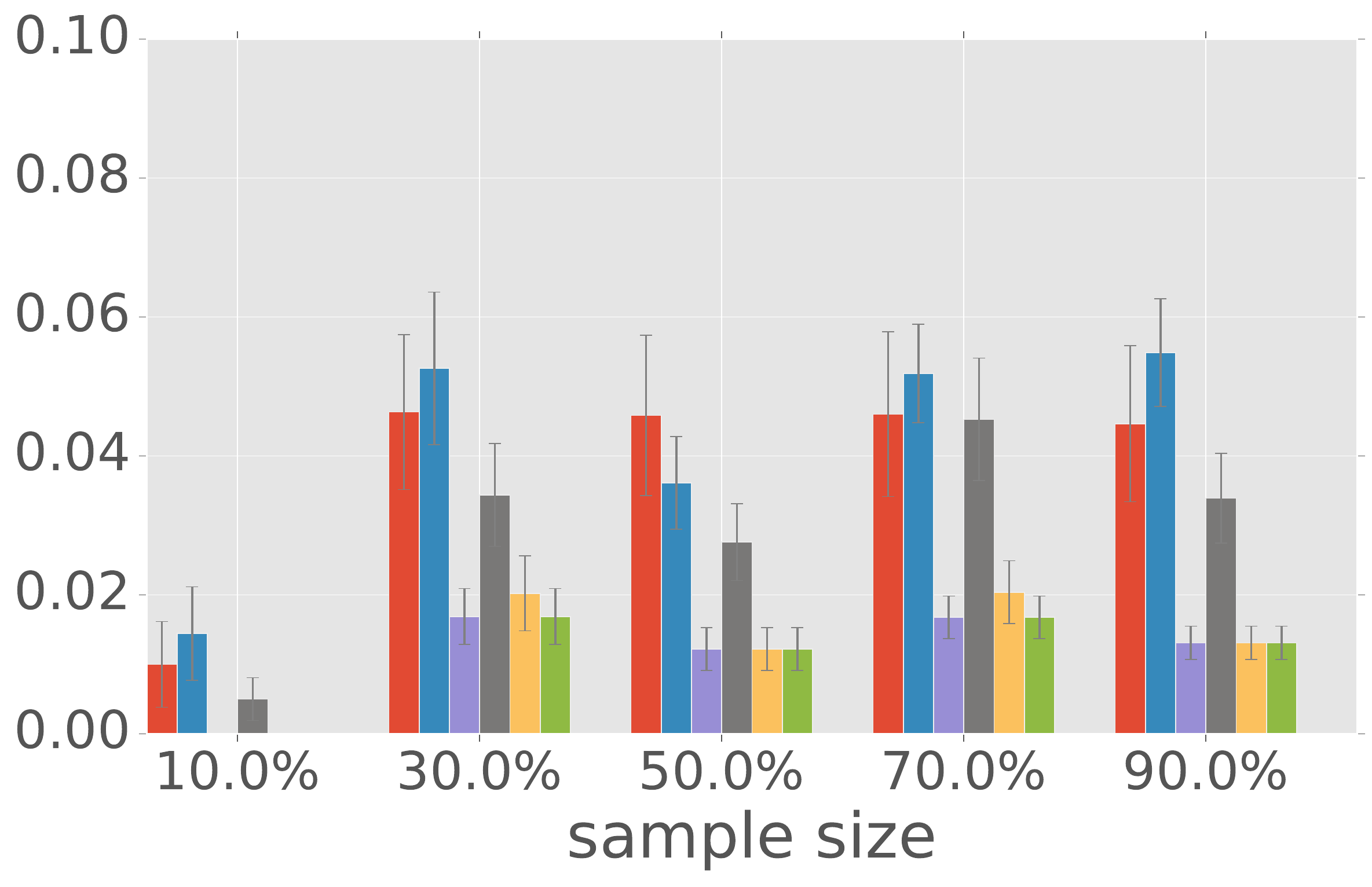}
        \label{fig:incremental-n20-75-null-2}
    }\hspace*{\fill}
	\subfloat[][\scriptsize 75\% Null: Avg. Power]{
    	\includegraphics[height=1.8cm]{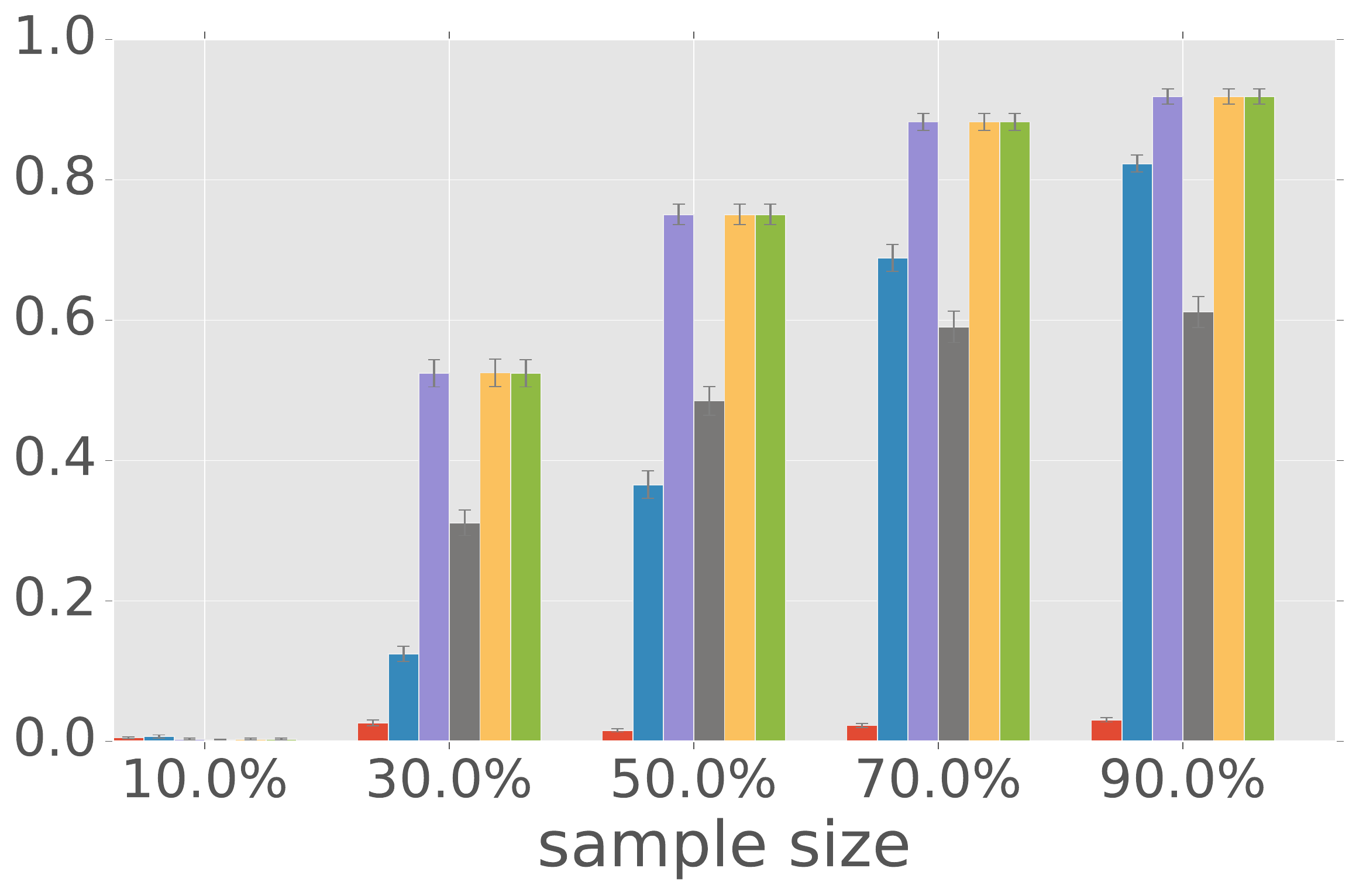}
    	\label{fig:incremental-n20-75-null-3}
    }
\\
    \subfloat{
    	\includegraphics[width=0.7\textwidth]{figures/exps/incremental/legend_incremental.pdf}
    }
    \caption{Exp.1c: Incremental Procedures on Synthetic Data / Varying Sample Size}
    \label{fig:exp-1c}
\end{figure*}

\subsection{Exp.1a: Static Procedures }
\label{sec:exp-1a}
In the first experiment we evaluate the static multiple hypothesis controlling procedures over synthetic data to motivate our choice of FDR (and similarly \emph{mFDR}) over FWER and per-comparison error rate (PCER) (i.e. no multiple hypothesis control). 

We created a large simulation study similar to the one in \cite{benjamini1995controlling} with $m$ hypotheses, ranging from 4-64.  Each hypothesis is comparing the expectations of two independently distributed normal random variables of variance 1 but different expectations varying from $5/4$ to $5$. The true null hypotheses are generated uniformly distributed across all tests and the proportions of true null hypotheses are set to $75\%$ and $100\%$ (i.e., completely random data).
We repeated the experiment 1,000 times. 

Figure~\ref{fig:exp-1a} shows the results for the static procedures, the Bonferroni-Correction (Bonferroni), the Benjamini-Hochberg procedure (BHFDR) and per-comparison error rate (PCER). For each procedure, we show the average number of discoveries, the average false discovery rate (FDR) and the average power.
Note that the power is 0 for all procedures over completely random data and thus, not shown. 

We observe that PCER has the highest power Figure~\ref{fig:exp-1a}(c), meaning that it can identify the highest proportion of truly significant discoveries.  However, PCER has also the highest false discovery rate across all configurations (see (b) and (e)). On completely random data, PCER averages 60\% false discoveries when testing 64 hypotheses in Figure~\ref{fig:exp-1a}(e).  Therefore PCER is not the right controlling target in multiple hypothesis testing in data exploration.

On the other hand, the Bonferroni procedure has the lowest average false discovery rate (see (b) and (e)), but the number of discoveries is also the lowest and the power also degrades quickly with an increasing number of hypotheses.  For this reason, FWER is too pessimistic for data exploration.

As a result, we advocate to use FDR (and similarly mFDR) as the control target for data exploration since we observed that the static FDR procedure, BHFDR, achieves a lower average error rate than PCER and and higher power than FWER. 



\subsection{Exp.1b: Incremental Procedures}
As discussed before it is not feasible to use the static procedures for interactive data exploration where the number of hypotheses are neither known upfront nor the \pvals can all be computed beforehand. 
For the remainder of the evaluation, we therefore focus on incremental procedures. 

Figure \ref{fig:exp-1b} uses the same setup as in Section \ref{sec:exp-1a}. The true null hypotheses are generated uniformly distributed across all tests and the proportions of true null hypotheses are set to $25\%$, $75\%$ and $100\%$ (i.e., completely random data).
In this experiment, we compare the different \ainv rules we developed, namely, $\beta$-farsighted with $\beta = 0.25$, $\gamma$-fixed with $\gamma = 10$, $\delta$-hopeful with $\delta = 10$, $\epsilon$-hybrid with $\epsilon = 0.5$, and $\psi$-support, against the non-interactive \sfdr (SeqFDR) procedure. 
The $\alpha$ for each procedure is set to $0.05$ and the $\epsilon$-hybrid uses unlimited window size.  
The $\psi$-support rule is implemented on top of $\gamma$-fixed.
We pre-set the values based on rule-of-thumb judgements and did not further tune them. 

Figure~\ref{fig:exp-1b}(b)(e)(h) show that all procedures control the FDR at level $\alpha=0.05$, barring some variation in the realization of the average FDR between the procedures (here lower is better).
Sequential FDR has the highest average FDR close to $0.05$, whereas the \ainv procedures on average make less mistakes. 
Next, we study the difference in FDR and the power of the \ainv rules, given different contexts of data exploration.

\begin{figure*}
\centering
\captionsetup[subfloat]{farskip=2pt,captionskip=1pt}
    \subfloat[][\scriptsize Census: Avg. Disc]{
  	 	\includegraphics[height=2.1cm, trim=5mm 0mm 4mm 0mm]{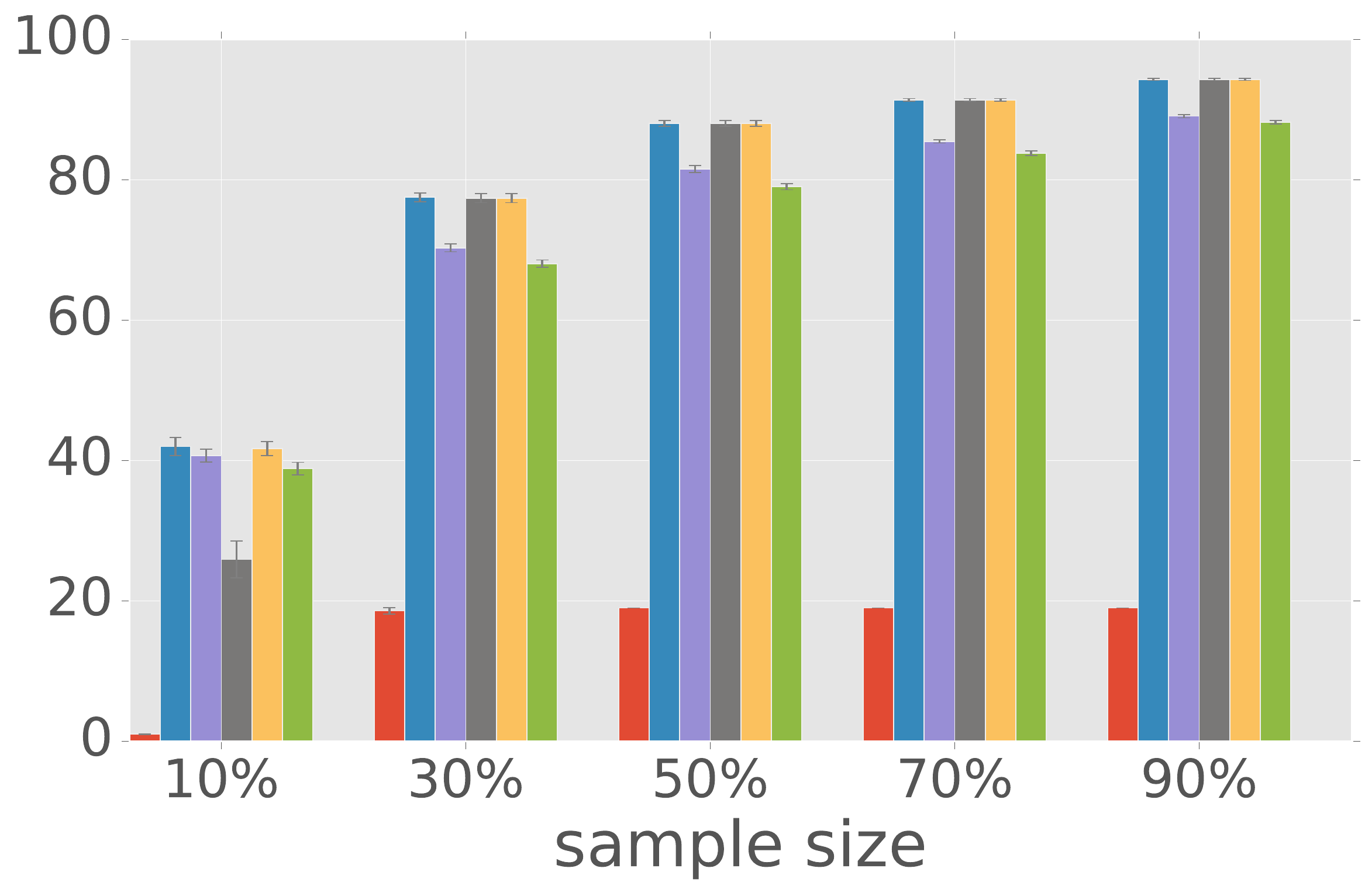}
	    \label{fig:incremental-census-1}
    }\hspace*{\fill}
    \subfloat[][\scriptsize Census: Avg. FDR]{
        \includegraphics[height=2.1cm, trim=5mm 0mm 4mm 0mm]{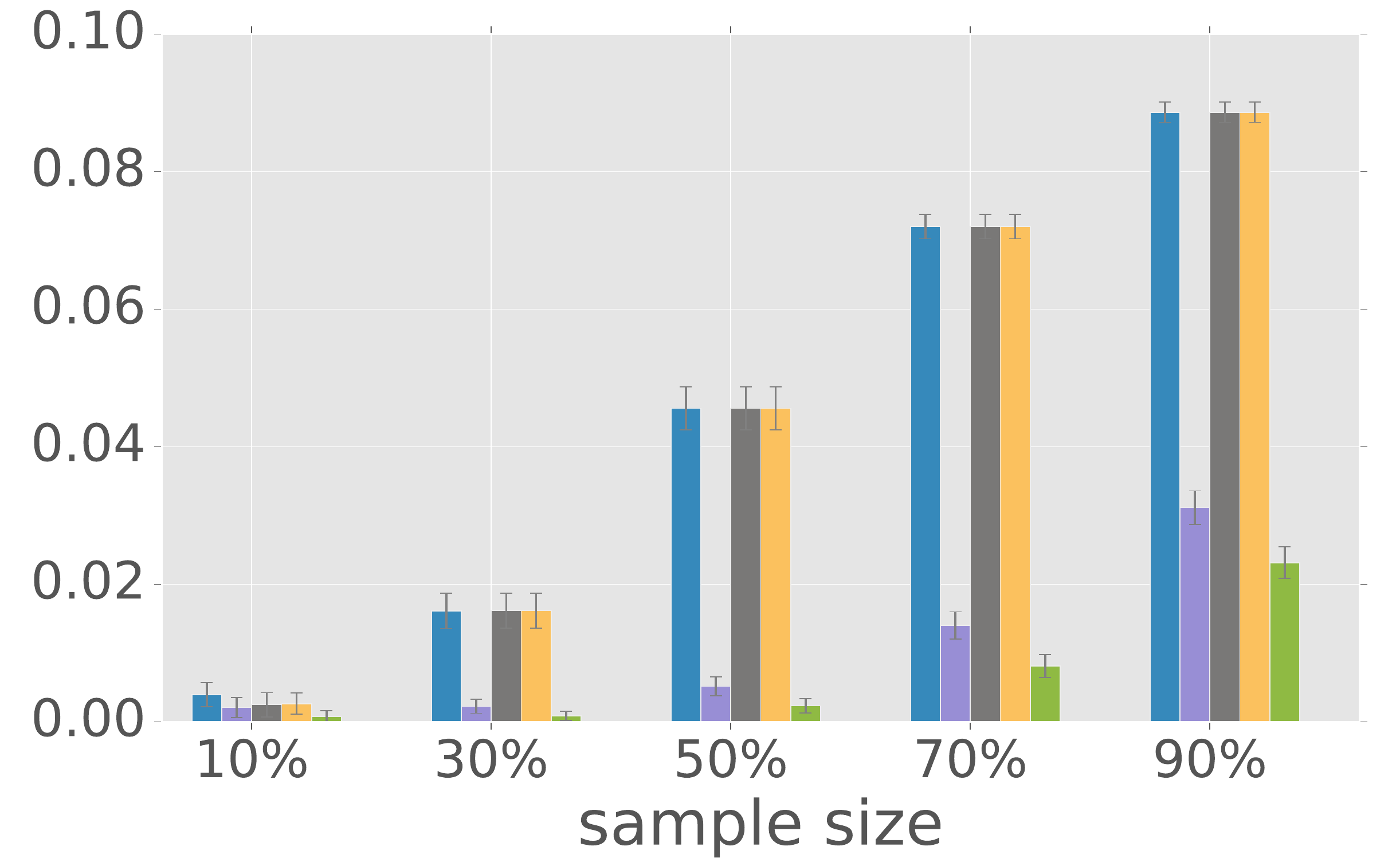}
        \label{fig:incremental-census-2}
    }\hspace*{\fill}
	\subfloat[][\scriptsize Census: Avg. Power]{
        \includegraphics[height=2.1cm, trim=5mm 0mm 4mm 0mm]{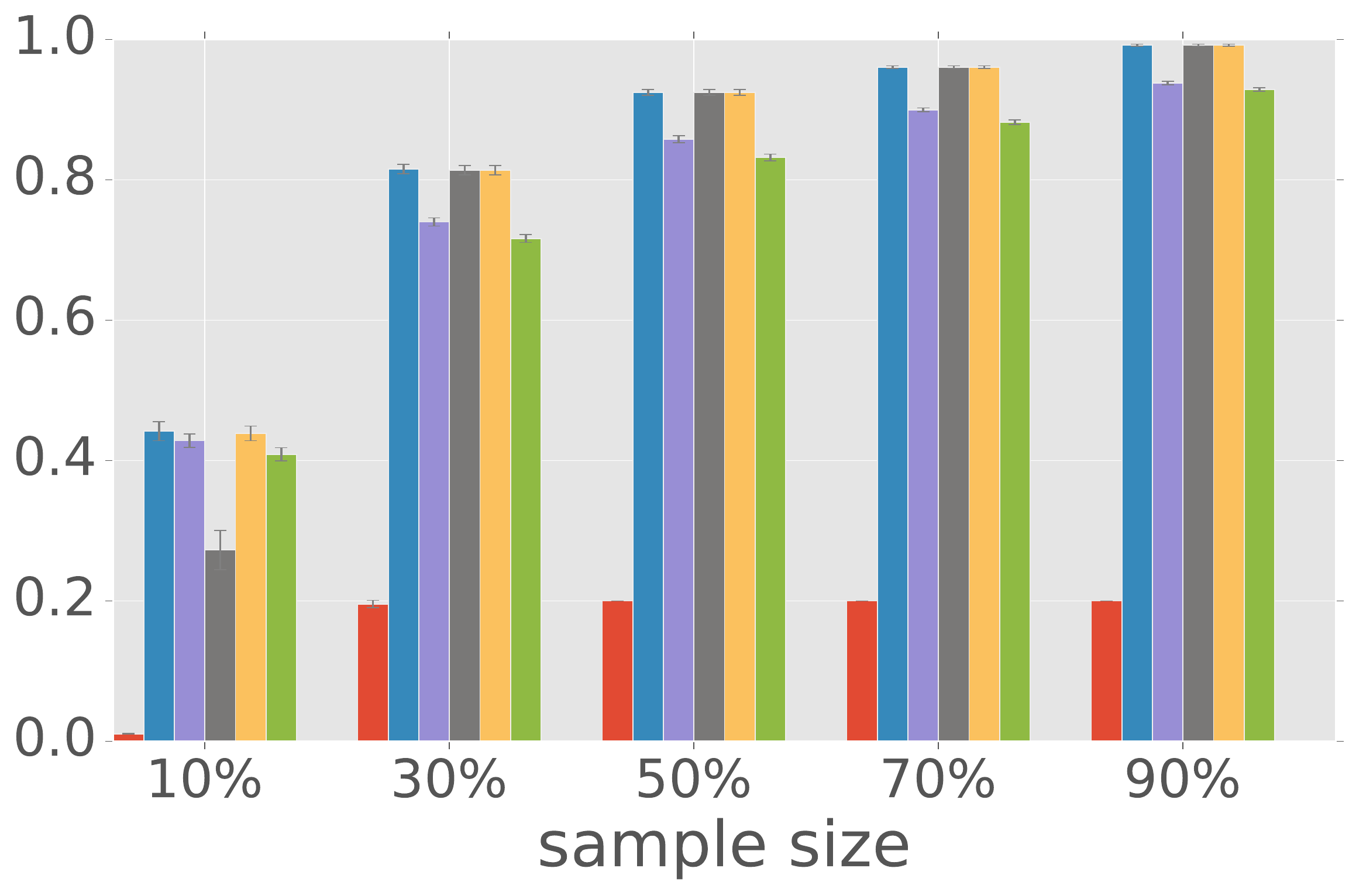}
        \label{fig:incremental-census-3}
     }\hspace*{\fill}
	\subfloat[][\scriptsize Rand. Census: Avg. Disc]{
        \includegraphics[height=2.1cm, trim=5mm 0mm 4mm 0mm]{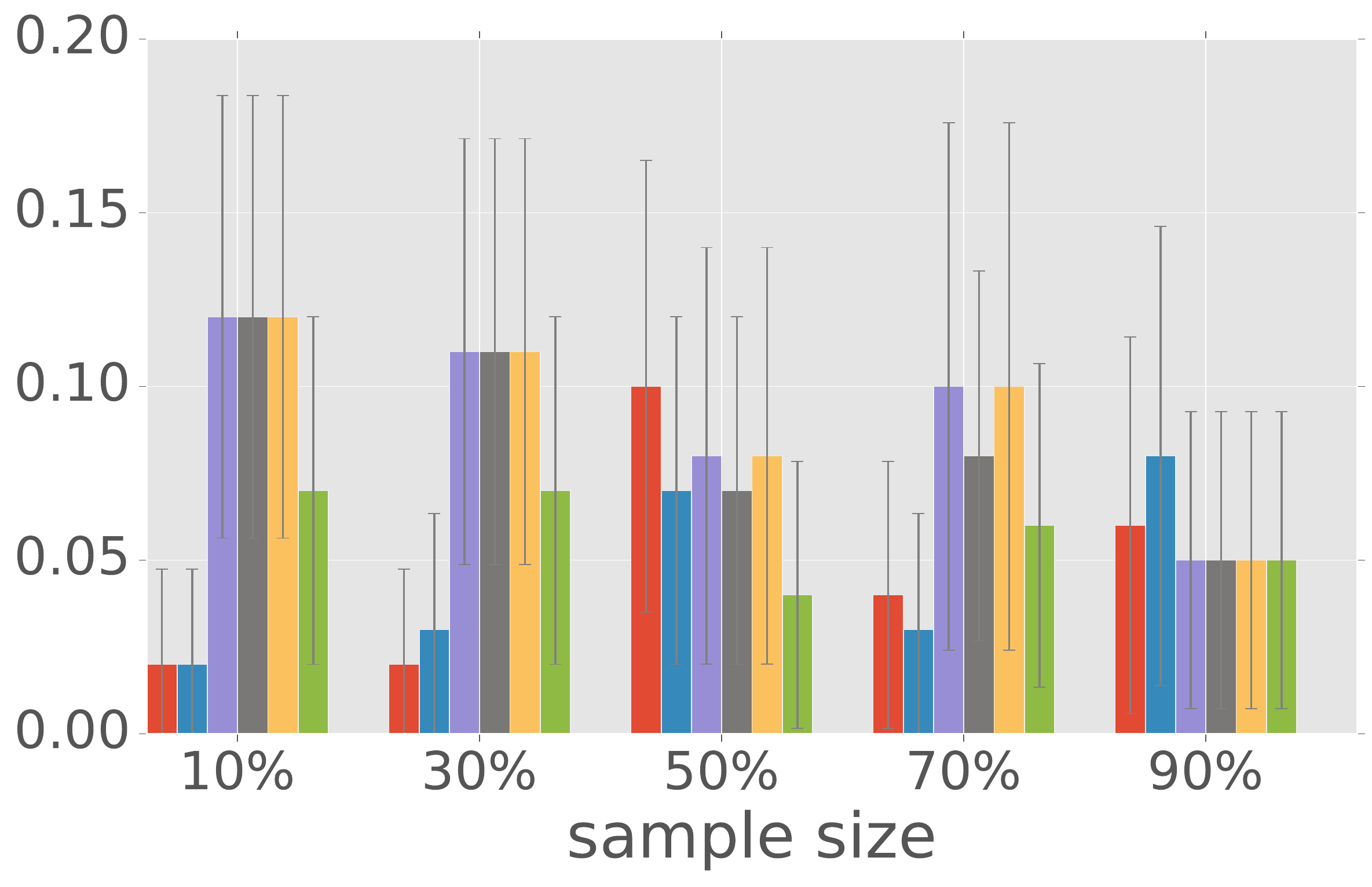}
        \label{fig:incremental-rcensus-1}
 	}\hspace*{\fill}
	\subfloat[][\scriptsize Rand. Census: Avg. FDR]{
        \includegraphics[height=2.1cm , trim=5mm 0mm 4mm 0mm]{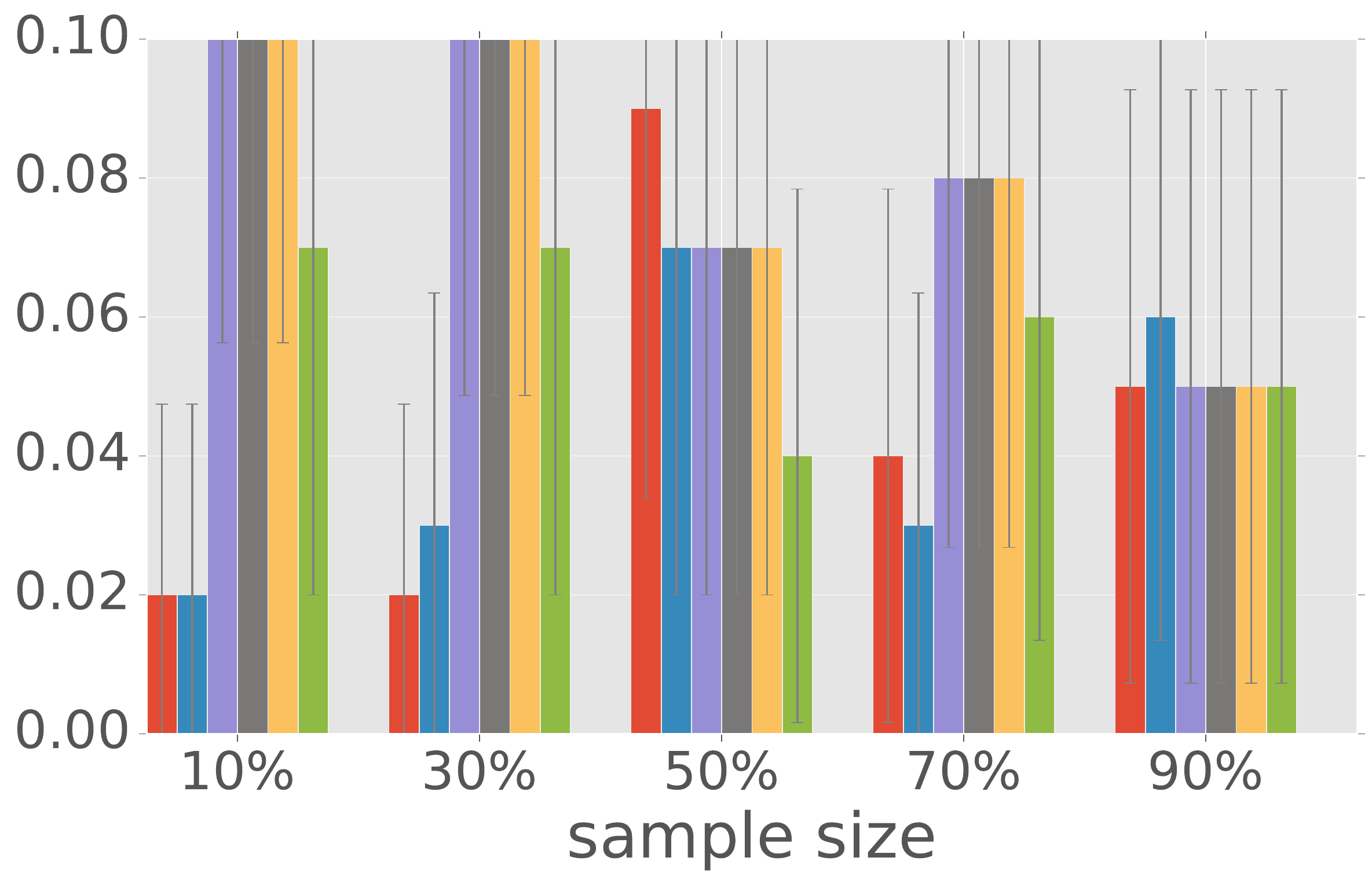}
        \label{fig:incremental-rcensus-2}
   }
   \\ 
   \subfloat{
        \includegraphics[width=0.7\textwidth]{figures/exps/incremental/legend_incremental.pdf}
    }
    \caption{Exp.2: Real Workflows on Census and Random Census Data}
    \label{fig:exp-2}
\end{figure*}

\subsubsection{Varying Number of Hypotheses}
With $\beta=0.25$, $\beta$-farsighted simulates a scenario in which the user is more confident or cares more about early discoveries being significant.  In this setting, $\beta$-farsighted is expected to make less significant discovers in a long run if the dataset has more randomness.  Figure \ref{fig:exp-1b}(f) shows that $\beta$-farsighted has very high power early on during the exploration, while it lowers gradually as more hypotheses are made. On the other hand, if the dataset has less randomness, such as in the \emph{25\% Null} configuration, $\beta$-farsighted is rewarded with the many discoveries during the exploration, and thus maintain its power for a longer run.

\subsubsection{Varying Degree of Randomness}
Figure\ref{fig:incremental-75-null-3} shows that when the data has more randomness, the $\gamma$-fixed rule tends to be more powerful than $\delta$-hopeful as the number of hypotheses increases.
When the data has less randomness, the $\delta$-hopeful rule becomes more powerful than $\gamma$-fixed rule. The reason is that the $\omega$ return from more frequent significant discoveries tends to keep the $\alpha$-wealth high, and since $\delta$-hopeful invests a fraction of the $\alpha$-wealth from the last rejected hypothesis, $\alpha$ per test tends to be high and hence the increase of power.

In light of this observation, we developed the previously mentioned $\epsilon$-hybrid that estimates the randomness in the dataset based on the history of hypothesis tests and picks between $\gamma$-fixed or $\delta$-hopeful. Figure \ref{fig:exp-1b} shows that $\epsilon$-hybrid procedure using $\epsilon=50\%$ of past rejections as the randomness threshold achieves overall a more robust performance in terms of power and FDR on varying degree of randomness than the aforementioned two procedures alone.
When the dataset is completely random, our $\alpha$-investing rules achieve similarly low false discovery rate as the Sequential FDR below 5\%. This provides the simulation-based evidence that our $\alpha$-investing rules correctly control the mFDR at $\alpha=5\%$.

Overall the results suggest that the performance of a given \ainv rule depends on how well its heuristic fits the context such as the importance of early discoveries and the data randomness. $\beta$-farsighted is suitable when the early hypotheses are more important than the later ones; whereas $\epsilon$-hybrid strategy provides more robust performance across varying degree of randomness. 

\subsubsection{Varying Support Size}
As part of interactive data exploration, the user usually applies various filter conditions, which change the support size for the different tests. 
To evaluate the impact of varying support sizes, we used the same setup as in Section \ref{sec:exp-1a}, but fixed the number of hypotheses to 64 and varied the sample size from $10$-$90\%$. 
The results are shown in Figure~\ref{fig:exp-1c}. 

While again $\epsilon$-hybrid and $\psi$-support do well across all configurations, $\psi$-support achieves lower average FDR especially for less random datasets (see Figure~\ref{fig:exp-1c}(b) and (e)). 
This is expected as the merit of the $\psi$-support rule is that it factors the support size of the hypothesis into the budget. Thus the rule tends to lower the per-test significance level when a low test \pval is observed on data of suspiciously low support size.

\subsection{Exp.2: Real Workflows}
In this experiment we show the effectiveness of our proposed procedures with real user workflows on the Census dataset \cite{census}.  
We collected the workflows of 115 hypothesis based on a user study we performed. 
The hypotheses were mostly formed by comparing histogram distributions by different filtering conditions, similar to the examples from Section~\ref{sec:requirements}.
We fixed the order of the hypotheses throughout the experiment as many of the hypotheses may depend on each other. 


To determine ground truth, we run the Bonferroni procedure with the user workflow on the full-size Census dataset to label the significant observations. We then down-sample the full data repetition for additional uncertainty.  Note that this evaluation method is a straw man as we do not know the actual truly significant observation on Census data.  It is likely to be biased towards towards more conservative $\alpha$-investing rules with more evenly distributed budgets, such as $\gamma$-fixed and $\psi$-support.


Figure~\ref{fig:exp-2}(a)-(c) shows the result of the user workflows over the Census data.
The $\gamma$-fixed and $\psi$-support rules perform better with average FDR significantly below $\alpha=0.05$, as shown in Figure~\ref{fig:exp-2}(b). For the other rules, the subtle side-effect of our label generation can be seen: the average false discovery rates for $\epsilon$-hybrid, $\beta$-farsighted and $\delta$-hopeful slightly inflate as the sample size increases, and reach over $\alpha=0.05$ to 0.09 for 90\% samples.
The reason is two-fold: 
First, the mFDR as the ratio of expectations is not necessarily bounded for only a particular fixed set of workflows. 
Second, the Bonferroni procedure generates a ground truth with a bias towards conservative \ainv rules with more evenly distributed budgets. Hence the more optimistic \ainv rules tend to make more mistakes.  This observation leads to interesting insight about the conservativeness of different \ainv rules.

To better demonstrate how our procedures control the false discovery rate, we therefore repeat the same experiment based on the real-world workflows but on randomized Census data. 
Figure~\ref{fig:exp-2}(d) and (e) show the results (note that the power for all procedures is by definition zero as all discoveries contribute to falsehood). 
We observe that the \ainv procedures remain comparable to the SeqFDR for higher sample sizes in terms of average FDR, although some variation exists such that some of the error rates have confidence intervals over the range 0.05 to 0.10.  We attribute this variation to the characteristic of our set of user-study workflows. 
For smaller sample sizes, we see higher variations.
We attribute this variation to the characteristic of our set of user-study workflows.

\section{Related Work}
\label{sec:related}
There has been surprisingly little work in controlling the number of false discoveries during data exploration even. 
This is especially astonishing as the same type of false discovery can also happen with traditional analytical SQL-queries. 
To our knowledge this is one of the first works trying to achieve a more automatic approach in tracking the user steps. 

Most related to this work are all the various statistical methods for significance testing and multiple hypothesis control.
Early works tried to improve the power of the Family Wide Error Rate using adaptive Bonferroni procedures such as S\v{i}d\'{a}k~\cite{vsidak1967rectangular}, Holm~\cite{holm1979simple}, Hochberg~\cite{hochberg1988sharper}, and Simes~\cite{simes1986improved}. However, all these methods lack power in large scale multi-comparison tests. 
 
The alternative False Discovery Rate measure was first proposed by Benjamini and Hochberg~\cite{benjamini1995controlling}, and soon became the statistical criteria of choice in the statical literature and in large scale data exploration analysis for genomic data ~\cite{mcdonald:statistics}. The original FDR method decides which hypotheses to reject only after all hypotheses were tested. Data exploration motivated the study of more advance techniques, such as sequential FDR~\cite{GSell} and $\alpha$-investing~\cite{foster2008alpha}, that work in a scenario where hypotheses arrive sequentially and the procedure needs to decide "on the fly" whether to accept or reject each of the hypotheses before testing the next one, while maintaining a bound on the FDR. Depending on the observed order of hypotheses, Sequential FDR can overturn previously accepted hypotheses into rejections based on the subsequent hypotheses. 
 
 $\alpha$-investing procedure also has revisiting policies that can potentially overturn previous decisions. The implication is that these procedures are incremental but non-interactive, because they require observing all the hypotheses before finalizing the decisions. However, it is often infeasible to obtain all the possible hypotheses a priori.  Therefore our work concerns $\alpha$-investing procedure with policies that are both incremental and interactive.
 In addition, none of the work addresses the issue on how to automatically integrate these techniques as part of an data exploration tool.

\section{Conclusion and Future Work}
\label{sec:concl}

In this paper we presented the first automatic approach to controlling the multiple hypothesis  problem during data exploration. 
We showed how the \system{} systems integrates user feedback and presented several multiple hypothesis control techniques based on $\alpha$-investing, which control \emph{mFDR}, and are especially suited for controlling the error for interactive data exploration sessions. 
Finally, our evaluation showed that the techniques are indeed capable of controlling the number of false discoveries using synthetic and real world datasets. 
However, a lot of work remains to be done from  creating and evaluating other types of default hypothesis over developing new testing procedures (e.g., for interactive Bayesian tests) to investigating techniques to recover from cases where the user runs out of wealth. 
Yet, we consider this work as an important first step towards more sustainable discoveries in a time where more data is analyzed than ever before.


\balance
\begin{scriptsize}
\bibliographystyle{abbrv}
\bibliography{bib}
\end{scriptsize}

\appendix

\section{Symbol table} \label{appendix:symbols}

The following table summarizes the important symbols and notations used in this paper.

\setlength{\textfloatsep}{1pt}
\begin{table}[h!]
\centering
\scriptsize
\begin{tabular}{| l | l |}
\hline
 $H$ & The set $\{H_1,\ldots,H_m\}$ of null hypothesis observed on the stream.\\
 \hline
 $\mathcal{H}$ & The set $\{\mathcal{H}_1,,\ldots,\mathcal{H}_m\}$ of corresponding ``\emph{alternative hypotheis}''.\\
 \hline
 $R$& The number of null hypothesis rejected by the testing procedure\\& (i.e., the discoveries).\\
 \hline
 $V$ & The number of erroneously rejected null hypothesis \\& (i.e., false discoveries, false positives, Type I errors).\\
 \hline
 $S$ & The number of correctly rejected null hypothesis\\& (i.e., true discoveries, true positives,).\\
 \hline
 $R(j)$ & The number of discoveries after $j$ hypothesis have been tested.\\
 \hline
 $V(j)$ & The number of false discoveries after $j$ hypothesis have been tested.\\
 \hline
 $S(j)$ & The number of false discoveries after $j$ hypothesis have been tested.\\
 $m$ & The number of hypothesis being tested. \\
 \hline
 $p_j$ & The \pval corresponding to the null hypothsis $H_j$.\\
 \hline
 $W(0)$ & Initial wealth for the \ainv procedures.\\
 \hline
 $W(j)$ & Wealth of the \ainv procedures after $j$ tests. \\
 \hline
 $\alpha$ & Significance level for the test with $\alpha\in (0,1)$.\\
 \hline
 $\eta$ & Bias in the denominator for $mFDR_{\eta}$.\\
 \hline
\end{tabular}
\caption{Notation Reference}
\label{tab:notation}
\end{table}

\end{document}